\documentclass[twocolumn,mathlines]{aastex631}

\usepackage{amsmath,epsfig,url}

\usepackage{hyperref}
\usepackage{graphicx,epstopdf,float,color}
\usepackage{rotating,multirow}
\usepackage{verbatim}

\shorttitle{Lightcurve manifold}
\shortauthors{Hemmati et al.}

\begin{document}


\title{Reducing the Dimensions of AGN Lightcurve Manifolds}
\author[0000-0002-1234-5678]{Shoubaneh Hemmati}
\affiliation{IPAC, California Institute of Technology, Pasadena, CA 91125, USA}
\author[0000-0002-2413-5976]{Jessica Krick}
\affiliation{IPAC, California Institute of Technology, Pasadena, CA 91125, USA}
\author[0000-0003-2686-9241]{Daniel Stern}
\affiliation{Jet Propulsion Laboratory, Pasadena, CA 91011, USA}
\author[0000-0002-1340-0543]{Vandana Desai}
\affiliation{IPAC, California Institute of Technology, Pasadena, CA 91125, USA}
\author[0000-0002-9382-9832]{Andreas Faisst}
\affiliation{IPAC, California Institute of Technology, Pasadena, CA 91125, USA}
\author[0009-0009-7840-931X]{Lucas Martín-García}
\affiliation{Universidad Rey Juan Carlos, Mostoles, Madrid 28933, Spain}
\author[0000-0002-8990-2101]{Varoujan Gorjian}
\affiliation{IPAC, California Institute of Technology, Pasadena, CA 91125, USA}
\author[0009-0006-3071-7143]{Aryana Haghjoo}
\affiliation{Department of Physics and Astronomy, University of California Riverside, Riverside, CA 92521, USA}
\author[0000-0002-3641-4366]{Farnik Nikakhtar}
\affiliation{Yale University, New Haven, CT 06511, USA}
\author[0000-0002-3031-5279]{Troy Raen}
\affiliation{IPAC, California Institute of Technology, Pasadena, CA 91125, USA}
\author[0009-0009-3048-9090]{Sogol Sanjaripour}
\affiliation{Department of Physics and Astronomy, University of California Riverside, Riverside, CA 92521, USA}
\author[0000-0002-3713-6337]{Brigitta M Sipőcz}
\affiliation{IPAC, California Institute of Technology, Pasadena, CA 91125, USA}
\author[0000-0003-4401-0430]{David Shupe}
\affiliation{IPAC, California Institute of Technology, Pasadena, CA 91125, USA}

\email{shemmati@caltech.edu}

\journalinfo{\textcopyright\ 2025. All rights reserved. Accepted to the Astrophysical Journal.}

\begin{abstract}
The Active Galactic Nuclei (AGN) glossary is vast and complex. Depending on selection method, observing wavelength, and brightness, AGNs are assigned distinct labels, yet the relationship between different selection methods and the diversity of time-domain behavior within and across classes remains difficult to characterize in a unified framework. Changing-look AGNs (CLAGNs), which transition between classifications over time, further complicate this picture. In this work, we learn a data-driven, low-dimensional representation of multi-wavelength photometric light curves of AGNs, in which the structure of the projected manifold correlates with AGN class and independent spectroscopic properties. Using the NASA Fornax Science Platform, we assemble light curves from ZTF, Pan-STARRS, Gaia, and WISE/NEOWISE for two samples: (1) a heterogeneous set of $\sim$2{,}000 AGNs spanning $z \lesssim 1$, including SDSS quasars, variability-selected sources, and CLAGNs; and (2) a homogeneous sample of $\sim$65{,}000 narrow-line AGNs at $z \approx 0.1$ with well-characterized optical emission-line measurements. Without using class labels during training, the learned manifolds organize variability-selected AGNs into coherent regions of the low-dimensional space, distinguish between turn-on and turn-off CLAGNs, and place tidal disruption events in distinct regions. Manifold coordinates correlate with key spectroscopic and host-galaxy properties---including stellar mass, [O~III] luminosity, and D$_n$(4000)---demonstrating that heterogeneous multi-band variability can be combined in a purely data-driven manner to recover correlations with independent physical diagnostics, without requiring explicit physical modeling. These results show that manifold learning offers a practical, assumption-light approach for integrating time-domain surveys and prioritizing spectroscopic follow-up.
\end{abstract}

\normalfont\normalsize

\section{Introduction}

Active galactic nuclei (AGNs) are among the most luminous and variable sources in the universe, powered by accretion onto supermassive black holes. In the unified model, the diversity of AGN observational classes is primarily explained by orientation effects and obscuration rather than intrinsic physical differences \citep[e.g.,][]{Antonucci1993, Urry1995, Netzer2015}. For example, Type-1 AGNs, with a direct view of the central engine, exhibit broad emission lines in their optical spectra, whereas Type-2 AGNs have these regions obscured by a dusty torus, showing only narrow emission lines. This orientation-based framework has been successful in explaining a wide range of observed features, including the presence of polarized broad lines in some Type-2 sources \citep{Antonucci1984}. However, it treats AGN classification as static and does not incorporate temporal variability or evolutionary changes within individual sources.

A defining property of AGNs is their variability across all wavelengths \citep{Ulrich1997}. Optical and UV continuum fluxes vary stochastically by tens of percent over timescales from days to years, and X-ray fluxes can change even more rapidly \citep[e.g.,][]{Koshida2014, McHardy2006}. These fluctuations reflect dynamic processes in the accretion flow and the immediate surroundings of the black hole \citep[e.g.,][]{MacLeod2010, Kelly2009, Burke2021}. Reverberation mapping exploits this variability to estimate the size of the broad-line region and infer black hole masses \citep[e.g.,][]{Peterson2001, Bentz2013}. In the optical, host galaxy starlight can dilute variability, whereas UV bands trace thermal disk fluctuations, and IR variability reflects reprocessed emission from the dusty torus \citep{Shappee2014, Cackett2021}. These differences highlight the multi-scale nature of AGN variability and motivate multi-band studies of time-domain behavior.

One class of AGNs that challenges static classification schemes is the changing-look AGNs (CLAGNs), which switch between Type-1 and Type-2 spectral states or undergo ``turn-on'' and ``turn-off'' episodes on observable timescales. These transitions, marked by the appearance or disappearance of broad emission lines and significant continuum variability, are generally attributed to changes in accretion rate or obscuration along the line of sight \citep[e.g.,][]{LaMassa2015, Ruan2016, Stern2018, Yang2018, Graham2020, Ricci2023}. The first spectroscopically confirmed CLAGN was reported by \citet{LaMassa2015}, revealing a decline in optical and X-ray flux accompanied by the loss of broad H$\alpha$ emission. Since then, time-domain spectroscopic surveys have significantly expanded the known population, with multi-epoch observations confirming changes in Balmer line profiles and continuum luminosity across a growing number of sources \citep[e.g.,][]{Graham2020, Green2022, Yang2023, Yang2025, Wang2025}. Mid-infrared and optical diagnostics have been used to disentangle intrinsic changes in accretion from variable obscuration, with current evidence often favoring accretion-driven mechanisms \citep[e.g.,][]{Yang2018, Ross2020}.

Machine learning (ML) techniques have recently been adopted for CLAGN discovery and characterization. Supervised models such as random forest classifiers have been trained on optical variability features to identify numerous candidates, many subsequently confirmed through spectroscopy \citep[e.g.,][]{Lopez2022}. Deep learning approaches have also been applied to capture complex light-curve structure and distinguish CLAGNs from other variable AGN or transient populations \citep[e.g.,][]{Sanchez2021}. More broadly, ML offers scalable approaches for organizing AGN variability and identifying candidate classes, though most existing efforts rely on individual surveys rather than integrating multi-wavelength time-domain data.

In this work, we use dimensionality reduction on AGN time-series data drawn from multiple photometric archives, accessed via the NASA Fornax Science Platform. We process heterogeneous light curves into a consistent multi-band representation and construct a high-dimensional description of time-domain photometric behavior. We then apply non-linear dimensionality reduction to organize sources by similarity in their multi-band variability. Manifold learning has proven effective in studies of stellar variability \citep[e.g.,][]{Sokolovsky2017}, AGN light curves \citep[e.g.,][]{Faisst2019, Pantoja2022}, and galaxy spectral energy distributions \citep[e.g.,][]{Masters2015, Hemmati2019a, Hemmati2019b, Sanjaripour2024, Sanjaripour2025}. By projecting AGN sub-populations into a reduced-dimensional space, groupings, transitions, and outliers in time-domain variability become more readily apparent, providing an interpretable framework for comparison across samples.

\begin{figure*}[ht]
\gridline{
  \fig{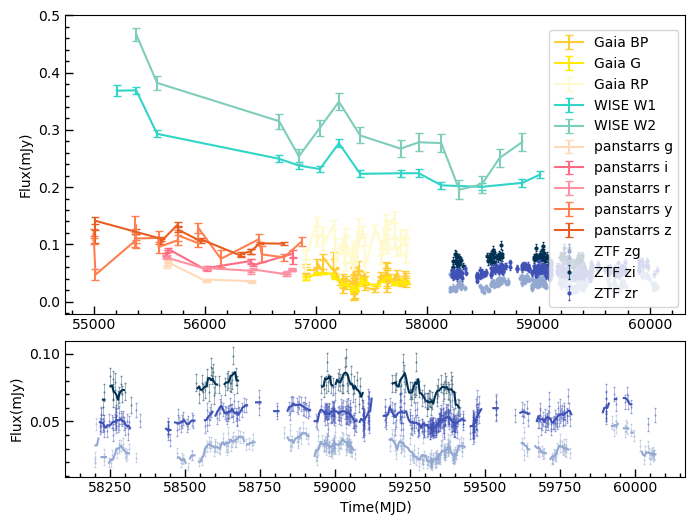}{0.48\textwidth}{}
  \fig{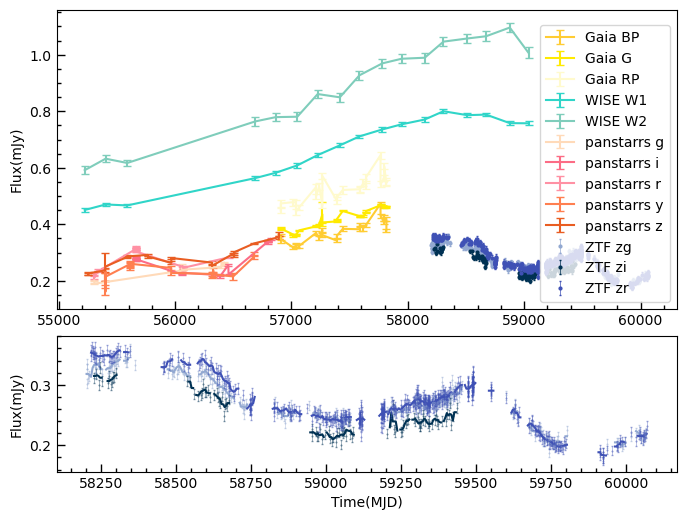}{0.48\textwidth}{}
}
\caption{Light curves of two AGNs at $z=0.277$ and $z=0.79$, queried from Gaia, ZTF, Pan-STARRS, and WISE, shown in the left and right panels, respectively. The bottom panels are zoomed-in views of the ZTF bands.}
\label{fig:sample_lightcurve}
\end{figure*}

Working with asynchronous, multi-band light curves remains a significant challenge in time-domain astronomy. Several recent efforts have developed statistically rigorous models for this task, including multivariate Gaussian Processes for modeling light curves across bands \citep{Gilbertson2020}, autoregressive models adapted for irregular time series \citep{Hu2020}, and Gaussian Process factor analysis for recovering latent variability structure \citep{Elorrieta2021}. These methods aim to model the underlying generative processes and perform interpolation under explicit statistical assumptions. In contrast, our approach uses observed light-curve behavior directly as input for dimensionality reduction, without attempting to model the generative variability process. Rather than focusing on prediction or imputation of missing data, our goal is to construct a similarity-based representation that captures dominant patterns in multi-band time-domain variability across diverse AGN populations.

We test this approach on two distinct AGN samples. The first is a heterogeneous compilation of AGNs spanning up to $z \sim 1$, including SDSS quasars, WISE- and GALEX-selected variable sources, and CLAGNs drawn from the literature. This sample is used to assess how labeled, variability-selected populations distribute within the learned time-domain space when labels are not used during training, testing whether population distinctions identified in individual surveys persist when all multi-band photometric light curves are analyzed jointly. The second sample is a homogeneous set of Type-2 AGNs at $z \sim 0.1$ with well-characterized optical emission-line diagnostics \citep{Kauffmann2003}. This sample provides extensive spectroscopic measurements, including [O~III] luminosity, H$\delta$ absorption, and D$_n$(4000), enabling a direct comparison between time-domain variability structure and independent host-galaxy and nuclear properties. Its uniform redshift coverage and clean classification make it a valuable reference set validating whether manifold coordinates learned from photometric variability alone exhibit systematic trends with independently measured spectroscopic and host-galaxy properties.

The remainder of this paper is organized as follows. In Section~\ref{sec:data}, we describe the time-series data and the Fornax platform. Preprocessing is presented in Section~\ref{sec:preprocessing}, and the two AGN samples are introduced in Section~\ref{sec:samples}. The manifold learning methodology is described in Section~\ref{sec:methods}, and results are presented in Section~\ref{sec:results}. We discuss limitations and implications and summarize our conclusions in Section~\ref{sec:conclusions}.

\section{Archival Time-Series Data}\label{sec:data}

As part of the NASA Fornax Initiative, we have developed and utilized the Fornax Science Platform to query and process time-series data from multiple NASA and non-NASA archives. Fornax is a cloud-based science platform designed to integrate data access, software environments, and scalable computing resources, enabling reproducible analysis of large, heterogeneous astronomical datasets. Throughout this work, Fornax is used as an access and processing framework rather than as a scientific method in itself. The Jupyter notebooks used to retrieve and preprocess the light curves analyzed in this paper are publicly available\footnote{See \textit{Fornax Lightcurve Notebooks} at \url{https://nasa-fornax.github.io/fornax-demo-notebooks/ml-agnzoo/}}.

This study incorporates time-domain photometric data from the following surveys: the Zwicky Transient Facility (ZTF; \citealt{ZTF}), the Wide-field Infrared Survey Explorer and its NEOWISE extension (WISE/NEOWISE; \citealt{WISE,Mainzer2011}), Pan-STARRS (Panoramic Survey Telescope and Rapid Response System; \citealt{Pan-STARRS}), and Gaia (\citealt{GAIA}). While additional time-domain datasets—such as those available through HEASARC (e.g., Fermi, BeppoSAX) or IceCube—could in principle be integrated into the same framework, we do not include them here due to differences in cadence, temporal baseline, wavelength regime, or sky coverage that would complicate direct comparison within the scope of this work.

ZTF is a wide-field, high-cadence optical survey conducted at the Palomar Observatory, covering approximately 25,000–30,000 square degrees of the northern sky in three filters ($g$, $r$, and $i$). For this analysis, we use the processed and cataloged DR18 photometry, accessed via positional matching with a $1''$ radius, rather than starting from image-level data. The ZTF data reduction and calibration pipeline is described in detail by \citet{Masci2019}.

Pan-STARRS operates a 1.8-m telescope in Hawaii and observes the sky north of declination $\sim -30^\circ$ in five filters ($g$, $r$, $i$, $z$, and $y$), with typically $\sim$12 epochs per filter. We retrieve Pan-STARRS photometry through cone searches with a $1''$ radius on the publicly released catalogs hosted on MAST. The survey design, data processing, and calibration procedures are described by \citet{Flewelling2020}.

Gaia is an ESA space mission that provides optical time-series photometry in three bands: a broad $G$ band and two narrower bands, $G_{BP}$ and $G_{RP}$. Gaia DR3 includes time-series data for approximately 11 million variable and non-variable sources. We retrieve Gaia light curves from the Gaia DR3 \textit{source\_lite} catalog \citep{Gaia2023} using \textit{astroquery}. The \textit{source\_lite} table provides a streamlined subset of the full \textit{gaia\_source} catalog optimized for large-scale cross-matching and light-curve access. We join uploaded target lists with the Gaia photometry tables to obtain multi-epoch flux measurements.

In the mid-infrared, WISE and NEOWISE provide all-sky coverage with a temporal baseline exceeding a decade and a typical cadence of $\sim$12 visits per sky location per epoch. We use the W1 (3.4 $\mu$m) and W2 (4.6 $\mu$m) bands from the WISE unTimely catalog \citep{Meisner2023}. The WISE photometry is accessed from publicly available AWS Open Data Repository buckets, stored in Parquet format and spatially partitioned using HEALPix indexing \citep{Gorski2005}, enabling efficient large-scale queries.

For each object, the queried photometry is stored in a Pandas multi-index DataFrame containing the object identifier, redshift (when available), an external label describing the object’s selection class (e.g., WISE-variable AGN or CLAGN), and the photometric measurements and uncertainties in each band. All fluxes are converted to consistent units ($\mu$Jy). The resulting dataset consists of asynchronous, irregularly sampled multi-band light curves with heterogeneous cadence and temporal coverage across surveys. Figure \ref{fig:sample_lightcurve} shows the lightcurves of two example quasars at $z=0.27$ and $z=0.79$. 

\begin{figure*}[ht]
\gridline{
  \fig{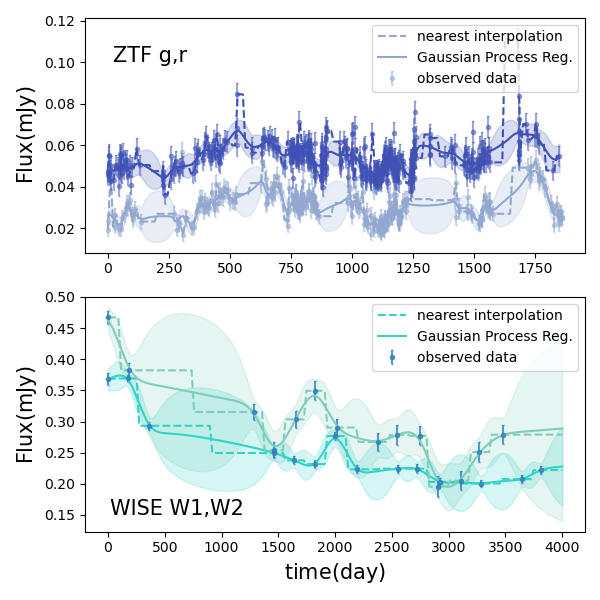}{0.48\textwidth}{}
  \fig{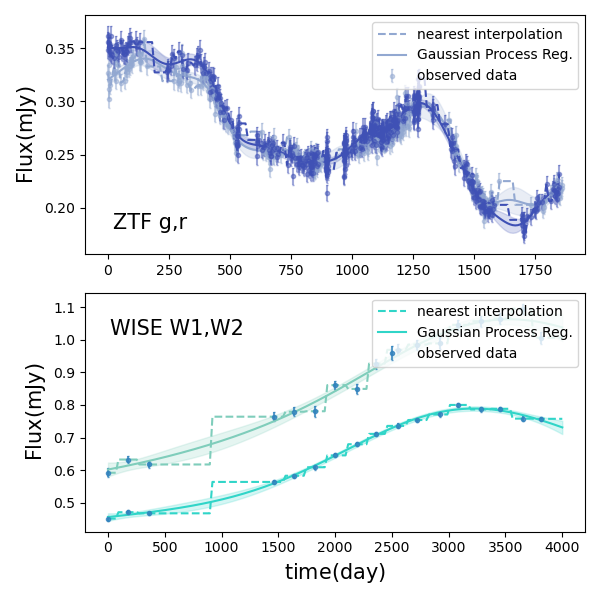}{0.48\textwidth}{}
}
\caption{Unified time arrays for the same two objects shown in Figure~\ref{fig:sample_lightcurve}. The ZTF $g$ and $r$ bands, along with the WISE W1 and W2 bands, are displayed. Gaussian Process Regression is shown with solid lines and shaded regions, and nearest-neighbor linear interpolation is depicted with dashed lines.}
\label{fig:interpolate_lightcurve}
\end{figure*}

\section{Preprocessing of Light Curves} \label{sec:preprocessing}

Although the light-curve data are stored in a consistent format and unit, additional preprocessing is required to enable a combined analysis across surveys and wavelength bands. Steps such as aligning time grids, handling missing data, choosing an effective temporal resolution, and applying flux normalization inevitably involve methodological choices that shape how variability is represented in the final analysis.

Because the light curves considered here are asynchronous and heterogeneous in cadence, depth, and temporal coverage across bands and surveys, many standard machine-learning methods cannot be applied directly. While several statistically rigorous approaches exist for modeling or imputing irregular time series (e.g., \citealt{Gilbertson2020, Hu2020, Elorrieta2021}), our goal at this stage is not to model the underlying generative variability process or to optimize interpolation performance. Instead, we adopt simple and transparent preprocessing strategies that are sufficient to place heterogeneous light curves on a common footing for similarity-based, distance-driven manifold learning. A comprehensive comparison to alternative imputation or time-series modeling frameworks is therefore beyond the scope of this work. For context, univariate time-series imputation methods have been extensively studied in the statistical literature (e.g., \citealt{Moritz2015, Beck2018}).

Most machine-learning algorithms operate on fixed-length inputs and cannot be applied directly to irregularly sampled light curves. Because our dataset combines multiple bands from multiple observatories, each with its own cadence, we unify the time grids by interpolating each band onto a common temporal grid. We explore two interpolation strategies: nearest-neighbor linear interpolation and Gaussian Process (GP) regression \citep{Williams1995} with a rational quadratic kernel. This kernel is chosen for its ability to represent variability across a range of characteristic timescales, effectively acting as a scale mixture of squared-exponential kernels. Figure~\ref{fig:interpolate_lightcurve} illustrates both interpolation methods for representative ZTF and WISE bands.

Nearest-neighbor interpolation provides a simple and transparent baseline that preserves observed values without introducing additional model assumptions. GP regression, in contrast, offers greater flexibility in modeling correlated variability and interpolating across gaps, while naturally incorporating measurement uncertainties through the covariance structure. GP regression is therefore well suited for heterogeneous astronomical time series, albeit at substantially higher computational cost. We retain both approaches in our initial analysis to illustrate sensitivity to interpolation strategy and to motivate our choice of GP-based unification for the remainder of the work.

Feature scaling is a standard preprocessing step for many machine-learning algorithms, particularly those based on distance metrics or gradient-based optimization. In the context of AGN light curves, the choice of normalization is non-trivial, as it should suppress trivial luminosity differences while preserving physically relevant variability patterns and relative flux information across bands. To this end, we normalize each object’s multi-band light curve by dividing all bands by the one-sigma-clipped maximum flux of the band with the largest number of observations. This choice ensures that the reference band is well sampled, reduces sensitivity to outliers, and preserves relative flux ratios between bands, thereby maintaining color information in the normalized light curves. We have verified that the qualitative structure of the learned manifold is robust to reasonable variations in normalization strategy, such as using a clipped median instead of a clipped maximum or selecting a different reference band, provided that relative flux structure across bands is preserved.

Finally, we apply a simple temporal alignment by shifting the start time of each light curve to zero (Figure~\ref{fig:interpolate_lightcurve}). This standardization allows light curves of different lengths and absolute epochs to be compared within a unified framework and is commonly used when the analysis focuses on variability patterns rather than absolute timing (e.g., \citealt{FOLGADO2018268}). We emphasize that this choice intentionally discards absolute timing information and is therefore not designed to capture inter-band time lags or reverberation effects. Investigating relative timing and lag structures in a multi-band framework is an important direction for future work but lies beyond the scope of the present analysis.

\begin{figure*}[ht]
    \centering
    \includegraphics[width=\textwidth]{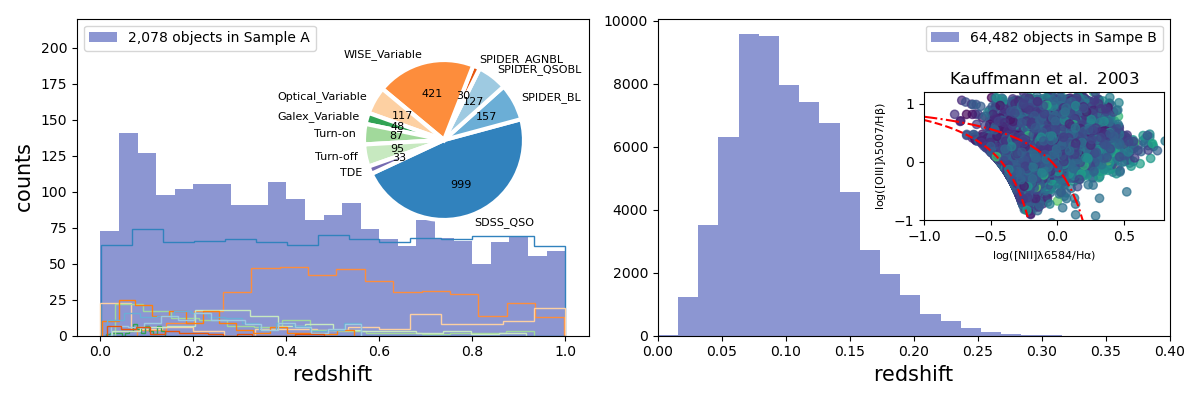}
    \caption{The redshift distribution of the two fully distinct samples is presented here. The left panel corresponds to first one with approximately 2,000 AGNs; the inset pie chart shows the number of objects from each sub-sample. The right panel comrises of approximately 65,000 Type-2 AGNs from \cite{Kauffmann2003}. The inset plot shows the sample on the BPT diagram, color-coded by the mean fractional variation of the lightcurves. All objects in this subsample sit in the composite and AGN parts of the BPT by construct.}
    \label{fig:sample}
\end{figure*}

\section{Two Distinct Samples from the Literature} \label{sec:samples}

To explore the structure of the multi-band AGN light curves in a data-driven manner, we compile two deliberately distinct samples from the literature, each designed to probe complementary aspects of AGN variability and population diversity. Sample~A is a heterogeneous compilation extending to $z\sim1$, assembled to test whether unsupervised analysis of multi-band light curves can recover known variability-selected populations, including CLAGNs and tidal disruption events (TDEs). Sample~B, in contrast, is a homogeneous, low-redshift ($z\sim0.1$) sample of narrow-line (Type-2) AGNs with uniformly measured spectroscopic and host-galaxy diagnostics, enabling a direct comparison between variability and independent physical observables. These two samples serve different purposes in the analysis. Sample~A is used to assess how labeled variability-selected populations project into the learned manifold when labels are not used during training. Sample~B is used to validate whether the manifold learned from photometric time series alone exhibits systematic trends with well-established spectroscopic and host-galaxy properties.

\subsection{Variables and CLAGNs (Sample A)}

The left panel of Figure~\ref{fig:sample} summarizes the components of Sample~A and their redshift distributions. This sample consists of AGNs drawn from multiple surveys and selection strategies, and individual objects may appear under more than one label. For example, an AGN may be both X-ray selected and identified as variable in WISE. Labels are therefore not mutually exclusive and are used solely for post-hoc interpretation of the manifold.

Approximately half of Sample~A is drawn from the SDSS DR16 catalog of spectroscopically confirmed quasars \citep{Lyke2020}, labeled as \texttt{SDSS\_QSO}. This subset is intended to serve as a broad reference population rather than a physically homogeneous class. Selection is therefore performed without imposing cuts on luminosity, black hole mass, host-galaxy properties, or variability metrics. To prevent this large parent catalog from dominating the learned manifold and to reduce unnecessary computational cost, we construct a stratified random subsample with an approximately flat redshift distribution over $0 < z < 1$. Specifically, the redshift range is divided into equal-width bins, and an equal number of quasars is randomly drawn from each bin, ensuring uniform redshift coverage while preserving diversity in intrinsic properties. Because the SDSS quasar catalog is large, many realizations of this stratified random sampling are possible. We therefore repeated the sampling procedure multiple times using different random draws and verified that the overall structure of the learned manifold and the qualitative relationships discussed below are stable across realizations. This subset intentionally overlaps with other labels in Sample~A, reflecting the fact that variability-selected AGN populations are not mutually exclusive.

We include AGNs from the SPIDERS (SPectroscopic IDentification of ERosita Sources) survey, which obtained targeted optical (BOSS) spectroscopy for X-ray–selected AGNs \citep{Clerc2016, Dwelly2017, Salvato2018}. Of the $\sim12{,}600$ successful SPIDERS spectra, roughly half are classified as quasars, $\sim4{,}800$ as galaxies, and the remainder as stars. We exclude stellar sources and restrict the sample to $0<z<1$. Broad-line AGNs and quasars are labeled as \texttt{SPIDERS\_AGNBL} and \texttt{SPIDERS\_QSOBL}, respectively, with their union denoted \texttt{SPIDERS\_BL}. Narrow-line AGNs are labeled \texttt{SPIDERS\_AGN}. This selection retains AGN-like extragalactic sources while excluding stars and purely star-forming systems.

Sample~A further includes variability-selected AGNs across multiple wavelengths. In the ultraviolet, we include 48 AGNs identified as variable in the GALEX time-domain survey \citep{Gezari2013} by \citet{Wasleske2022}, labeled \texttt{GALEX\_Variable}. Optical variables (\texttt{Optical\_Variable}) consist of 117 AGNs identified in the three-year COSMOS dataset from the VLT Survey Telescope \citep{Capaccioli2011} by \citet{DeCicco2019}. In the mid-infrared, we include WISE-variable AGNs selected from the R90 catalog of \citet{Assef2018}. Of the $\sim4.5$ million WISE-detected AGNs in this catalog, \citet{Prakash2019} identified 687 highly variable sources, of which 421 fall within our adopted redshift range and are labeled \texttt{WISE\_Variable}. No additional cuts are applied to these variability-selected subsamples beyond the original selection criteria and redshift range.

We additionally compile 182 changing-look AGNs from the literature \citep{LaMassa2015, MacLeod2016, Ruan2016, Yang2018, Sheng2020, Graham2020, Green2022, Lopez2022, Hon2022}. These are labeled as 87 Turn-On and 95 Turn-Off CLAGNs, based on the reported direction of their spectral transitions. The selection criteria for CLAGNs vary substantially across studies, ranging from large-amplitude optical variability thresholds \citep[e.g.,][]{MacLeod2016} to spectroscopically confirmed changes in Balmer line fluxes or widths \citep[e.g.,][]{Graham2020, Green2022}. Other works employ mid-infrared variability \citep{Sheng2020} or machine-learning–assisted photometric selection followed by spectroscopic confirmation \citep{Lopez2022, Hon2022}.

Given this heterogeneity, we do not attempt to homogenize CLAGN selection or assess the reliability of individual classifications. Instead, we group all confirmed or high-confidence sources into Turn-On and Turn-Off categories and treat the labels as external information. Our aim is to test whether unsupervised manifold learning applied solely to light-curve morphology can recover or distinguish these labeled transitions in variability space. Objects without clearly defined spectral transitions or ambiguous turn-on/off directionality were excluded.

Finally, we include 33 confirmed tidal disruption events (TDEs) from ZTF \citep{vanVelzen2021, Hammerstein2023, Somalwar2023}. TDEs are included because they exhibit extreme variability and spectroscopic changes that can resemble those seen in CLAGNs, offering a useful comparison class. We note that most TDE searches explicitly exclude known AGNs or sources with AGN-like mid-infrared colors to reduce contamination \citep[e.g.,][]{Stern2005, vanVelzen2021}, implying that TDEs occurring in active galaxies may be underrepresented despite being astrophysically expected \citep[e.g.,][]{Chan2019, Ricci2020, Neustadt2020, Wang2024}.

\subsection{Homogeneous Narrow-Line AGNs (Sample B)}

Sample~B is drawn from the Type-2 AGN catalog of \citet{Kauffmann2003} and has a narrow redshift distribution centered at $z\sim0.1$, as shown in the right panel of Figure~\ref{fig:sample}. These AGNs are selected based on their positions above the star-forming sequence in the composite and AGN regions of the BPT diagram \citep{BPT1981} and have well-characterized host-galaxy and emission-line properties.

This sample provides a complementary counterpart to Sample~A. Its uniform selection, large size, and consistent spectroscopic measurements enable a direct comparison between the variability manifold and independent physical diagnostics, including [O\,\textsc{iii}]$\lambda5007$ luminosity, stellar mass, D$_n$(4000), and H$\delta_A$. These quantities are not used in constructing the manifold and therefore offer an external validation of whether photometric variability morphology encodes physically relevant information.

We note that Sample~B is dominated by relatively low-luminosity AGNs, including Seyferts and LINERs, and does not fully overlap in luminosity with the more powerful quasars in Sample~A. Extending this analysis to higher-luminosity obscured quasars—for example, using optically selected Type-2 quasar samples from SDSS \citep{Zakamska2003, Reyes2008}—is a natural avenue for future work and would allow testing whether the variability-manifold structure persists across a broader range of accretion power.

\begin{figure*}[ht]
    \centering
    \includegraphics[width=\textwidth]{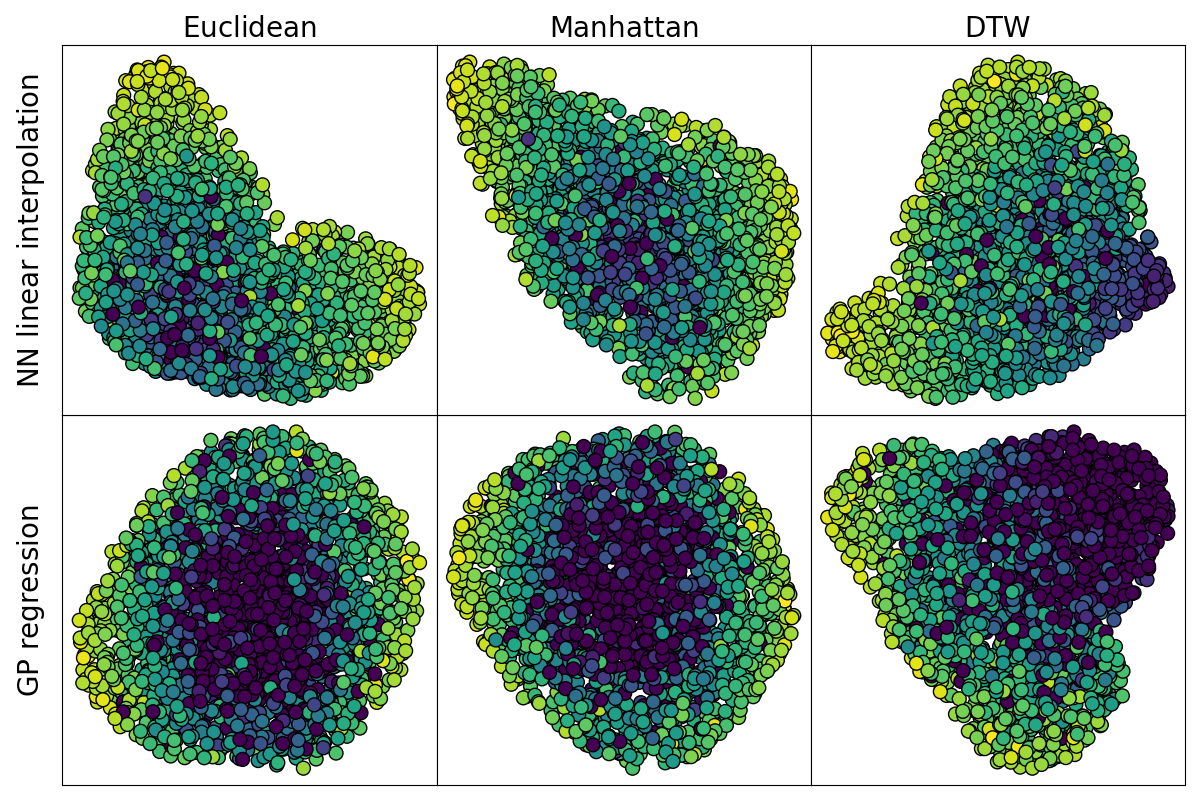}
    \caption{UMAP projections of the WISE W1 light curves for sample A, generated using different combinations of distance metrics (columns) and interpolation methods (rows). This sample includes AGNs compiled from archival variability catalogs, including changing-look AGNs (CLAGNs) and tidal disruption events (TDEs).  Each point represents a source, color-coded by the mean fractional variation of its light curve. The color scale ranges from dark blue (low variability) to yellow (high variability).These projections allow us to assess how different preprocessing choices affect the learned manifold structure. Among the configurations, Gaussian Process (GP) interpolation combined with the Dynamic Time Warping (DTW) distance metric produces the most coherent structure, with a clear gradient in fractional variation. The presence of this gradient indicates that variability amplitude is a key factor shaping the manifold, suggesting that this configuration best preserves physically meaningful variability information.
}
    \label{fig:umapparams}
\end{figure*}

\begin{figure*}[ht]
    \centering
    \includegraphics[width=\textwidth]{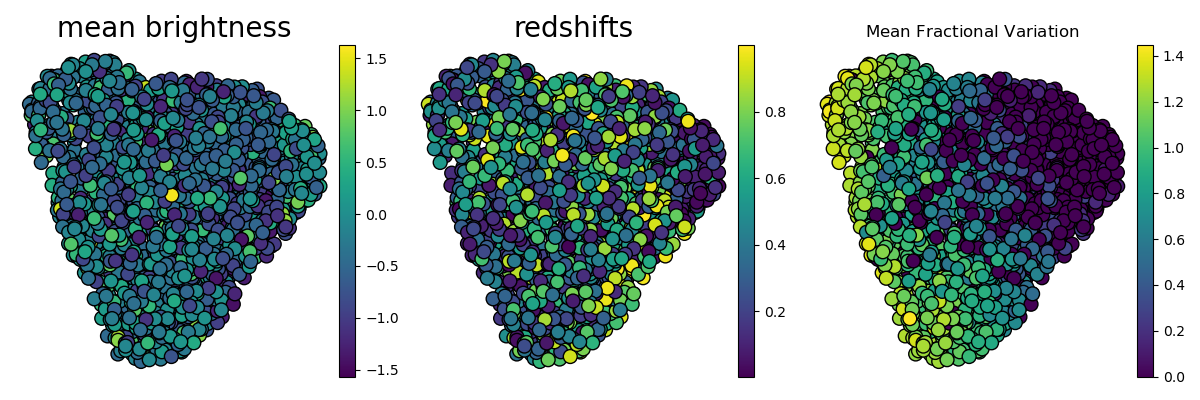}
    \caption{This figure presents UMAP projections of the WISE W1 light curves for the first AGN sample, using GP regression and the DTW distance metric (as in the bottom-right panel of Figure 4). Points are color-coded by mean brightness, redshift, and mean fractional variation (left to right). As expected, given that the manifold was constructed from normalized lightcurves, there is no clear structure associated with redshift or brightness. In contrast, a pronounced gradient is observed with mean fractional variation, indicating that variability amplitude plays a dominant role in shaping the learned manifold. This confirms that the embedding captures physically meaningful differences in AGN variability behavior.}
    \label{fig:umapw1-main}
\end{figure*}

\begin{figure*}[htb!]
    \centering
    \includegraphics[width=\textwidth]{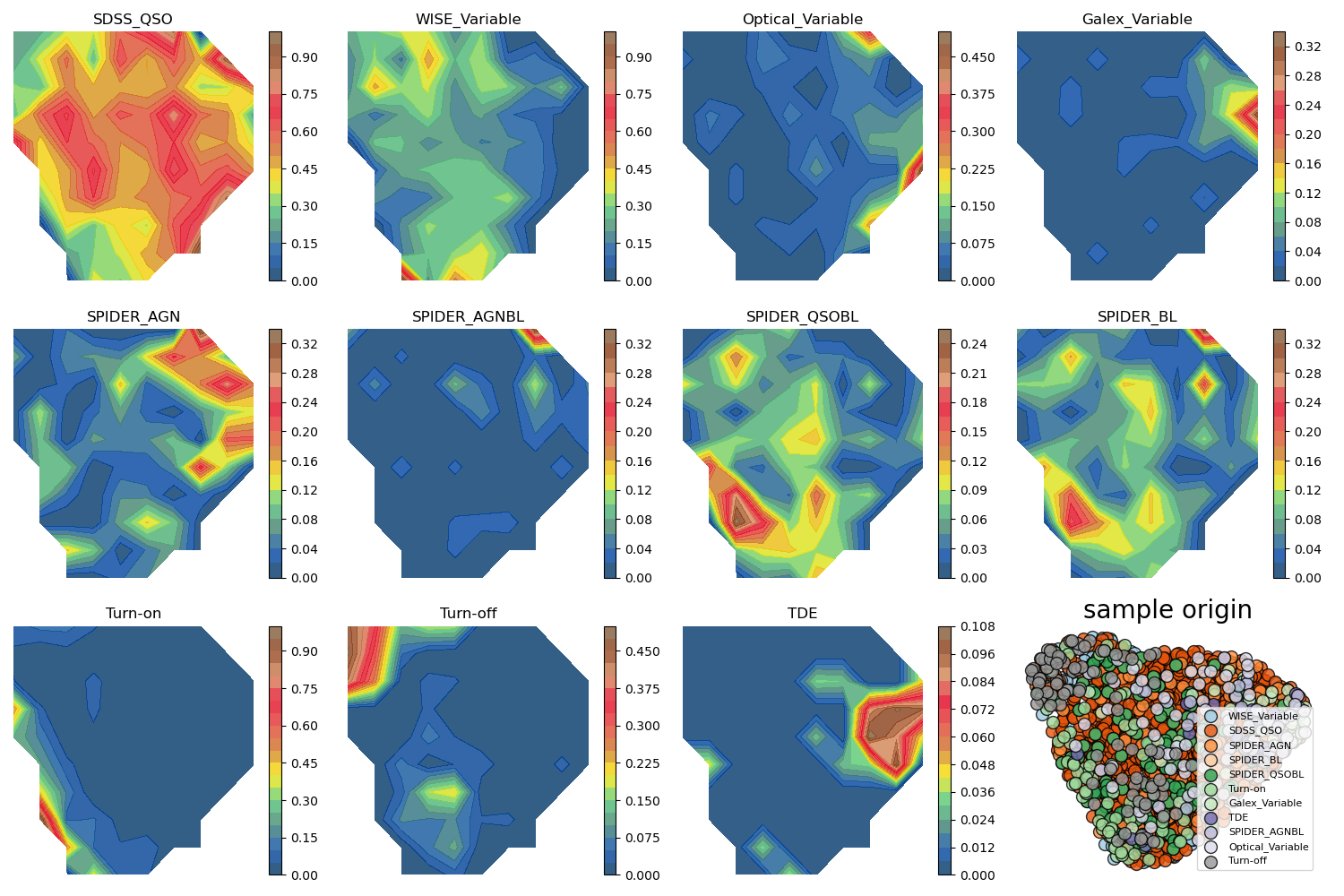}
    \caption{Separation of different labels in the first sample is illustrated using UMAPs, with parameters similar to those in Figure \ref{fig:umapw1-main}. The bottom right panel displays the origin labels for all objects in this sample. Each panel shows a 2D histogram over the UMAP manifold for a particular AGN class label. The color bar indicates the local point density for that class in the UMAP embedding space, with lighter shades representing regions of higher density. This allows for a visual comparison of where each labeled population is concentrated or dispersed within the learned manifold.  The degree of separation between these distributions indicates how well the manifold distinguishes between different AGN classes based on variability. These projections use the same preprocessing and UMAP parameters as in Figure 5, namely GP-regressed WISE W1 light curves and the DTW distance metric.}
    \label{fig:umapw1-sample}
\end{figure*}

\section{Manifold Learning and Dimensionality Reduction}\label{sec:methods}

In this work, we use Uniform Manifold Approximation and Projection (UMAP; \citealt{McInnes2018}) as a nonlinear dimensionality reduction technique to construct low-dimensional representations of multi-band AGN light curves. UMAP is designed to preserve local neighborhood structure in high-dimensional data while providing a computationally efficient embedding suitable for visualization and exploratory analysis. Throughout this work, UMAP is used to examine similarity structure in multi-band time-domain photometric behavior. While the resulting low-dimensional representations are not intended to provide precise metric distances or formal hypothesis tests, they offer an interpretable framework for identifying clustering, gradients, and transitions in variability space.

Briefly, UMAP constructs a weighted $k$-nearest-neighbor graph from the input data using a chosen distance metric, encoding local similarity relationships between points. This graph is interpreted as a fuzzy topological representation of the data under the assumption that it is sampled from a Riemannian manifold. An optimization procedure then seeks a low-dimensional embedding that preserves these local relationships as faithfully as possible. The resulting projection emphasizes relative similarity structure rather than absolute metric distances, and is therefore well suited for comparing multi-band time-domain variability patterns across heterogeneous datasets.

When configuring UMAP, three parameters play a primary role: the number of neighbors, the minimum distance, and the distance metric. The number of neighbors controls the scale over which local structure is defined. Larger values incorporate information from broader neighborhoods and tend to emphasize more global trends in the data, while smaller values prioritize fine-scale structure. In this work, we adopt relatively large neighborhood sizes to suppress stochastic noise and focus on broad variability similarities rather than object-level idiosyncrasies. The minimum distance parameter governs how tightly points are allowed to cluster in the low-dimensional space; we use a comparatively large value to discourage overly compact clusters and instead highlight continuous distributions and separations between populations.

The choice of distance metric is particularly important because it determines how similarity between light curves is quantified. While Euclidean distance is commonly used and performs well for uniformly sampled, low-dimensional data, it is less well suited for irregular, asynchronous time series. Manhattan distance can mitigate some high-dimensional effects, but still treats time series as static vectors. For this reason, we also explore Dynamic Time Warping (DTW), a distance metric specifically designed for time-series comparison.

DTW computes the similarity between two sequences by identifying an optimal alignment that allows local stretching or compression in time, thereby accommodating phase offsets and irregular sampling. This flexibility makes DTW well suited for AGN light curves, where variability timescales, cadences, and phase relationships can differ significantly across bands and objects. DTW has been widely used in temporal pattern recognition across many domains, including speech processing, gesture recognition, and biological signal analysis (e.g., \citealt{Keogh2005}). Its main limitation is computational cost, particularly for multivariate time series, since it requires evaluation of a full alignment cost matrix.

To assess the impact of preprocessing and distance choices, Figure~\ref{fig:umapparams} presents UMAP projections for Sample~A using three distance metrics (Euclidean, Manhattan, and DTW) applied to light curves interpolated via both nearest-neighbor linear interpolation and Gaussian Process (GP) regression. These projections are colored by the arctangent of the excess variance, a scalar summary of mean fractional variability amplitude. The arctangent transformation maps the non-negative excess variance values onto a bounded range, reducing the influence of extreme outliers while preserving relative ordering. Dark blue corresponds to lower variability amplitude and yellow to higher variability amplitude.

Among the tested configurations, DTW applied to GP-regressed light curves yields one of the clearest embeddings in the sense that sources with similar variability amplitudes populate contiguous regions of the projection and form smooth gradients. Other combinations—such as GP regression with Euclidean distance—also recover interpretable structure and are not excluded from consideration. We therefore emphasize that DTW+GP is not uniquely optimal, but represents one effective choice for the subsequent analysis. Figure~\ref{fig:umapparams} is included to demonstrate the sensitivity of the embedding to preprocessing and distance choices, rather than to define a single preferred configuration.

To aid interpretation of the embeddings, we color-code UMAP projections using three scalar quantities: redshift, mean brightness, and mean fractional variability. This allows us to assess whether the learned manifold organizes sources in a manner consistent with known observational properties, without using these quantities as inputs to the embedding.

The brightness parameter shown in the UMAP plots is defined as
\begin{equation}
F_{\text{mean}} = \log\left( \sum_{b=1}^{M} \langle F_b \rangle_{\sigma} \right),
\end{equation}
where the sum runs over the $M$ photometric bands, $F_b$ denotes the flux values in the $b^{\text{th}}$ band, and $\langle F_b \rangle_{\sigma}$ is the time-averaged flux after applying a 5$\sigma$ clipping to suppress outliers.

The mean fractional variability is defined as
\begin{equation}
\mathrm{Fvar}_{\text{mean}} = \arctan\left( \sum_{b=1}^{M} \frac{\sqrt{\sigma_b^2 - \delta_b^2}}{\langle F_b \rangle} \right),
\end{equation}
where
\begin{align}
\sigma_b^2 &= \frac{1}{N} \sum_{i=1}^{N} (F_i - \langle F \rangle)^2, \\
\delta_b^2 &= \frac{1}{N} \sum_{i=1}^{N} (\Delta F_i)^2,
\end{align}
and $\Delta F_i$ denotes the measurement uncertainty. This formulation accounts for observational noise following the standard excess-variance definition \citep{Peterson2001}. When the estimated intrinsic variance is smaller than the mean squared uncertainty, the contribution is set to zero, ensuring that $\mathrm{Fvar}_{\text{mean}}$ remains non-negative. The logarithmic and arctangent transformations are applied solely for visualization purposes, to compress dynamic range and enhance interpretability.

\begin{figure*}[ht]
    \centering
    \includegraphics[width=\textwidth]{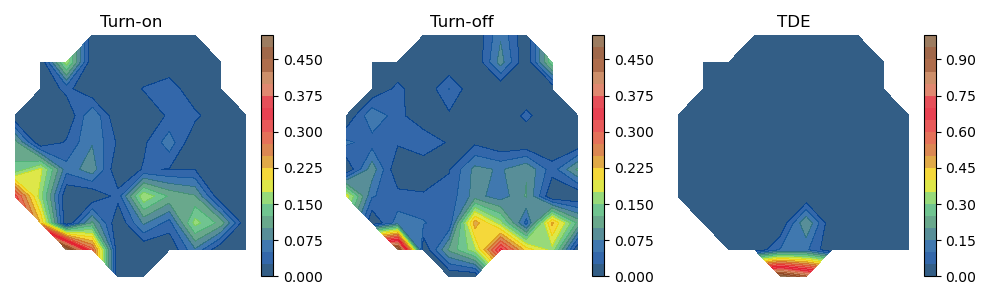}
    \caption{Similar to Figure \ref{fig:umapw1-sample} but trained on the ZTF g band alone to show the similarity of TDEs(right panel) to CLAGNs (two left panels) in optical.}
    \label{fig:umapzg-sample}
\end{figure*}

\begin{figure*}[ht]
    \centering
    \includegraphics[width=\textwidth]{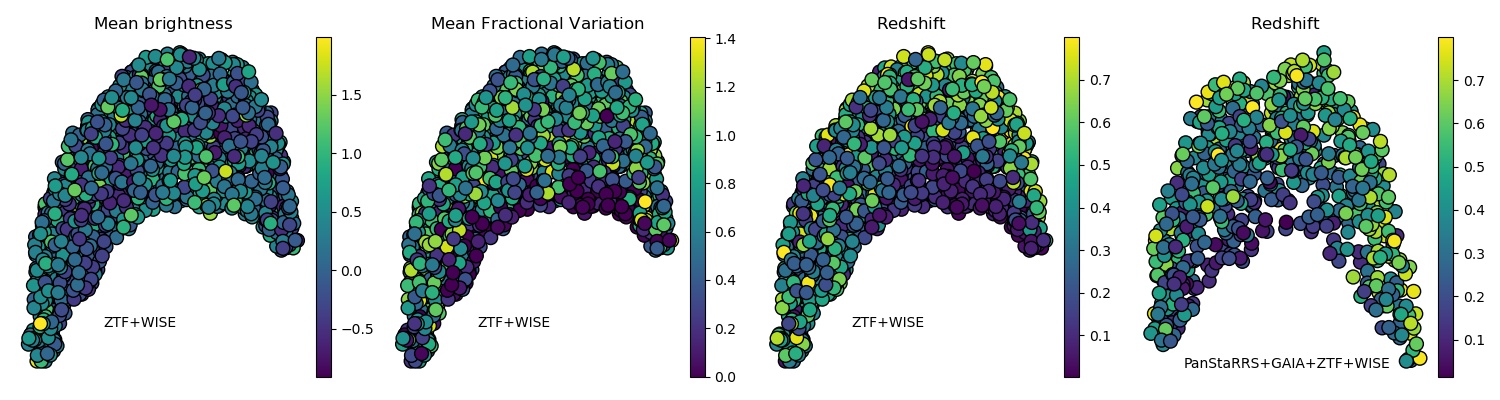}
    \caption{Similar to Figure \ref{fig:umapw1-main}, with the three left panels generated using the combined ZTF+WISE bands and the right panel from Pan-STARRS+Gaia+ZTF+WISE. With the increased number of wavebands covering a broader range of the SED, a redshift trend becomes apparent, and the trend with mean fractional variation is less prominent.}
    \label{fig:umapztfw-main}
\end{figure*}

\begin{figure*}[ht]
    \centering
    \includegraphics[width=\textwidth]{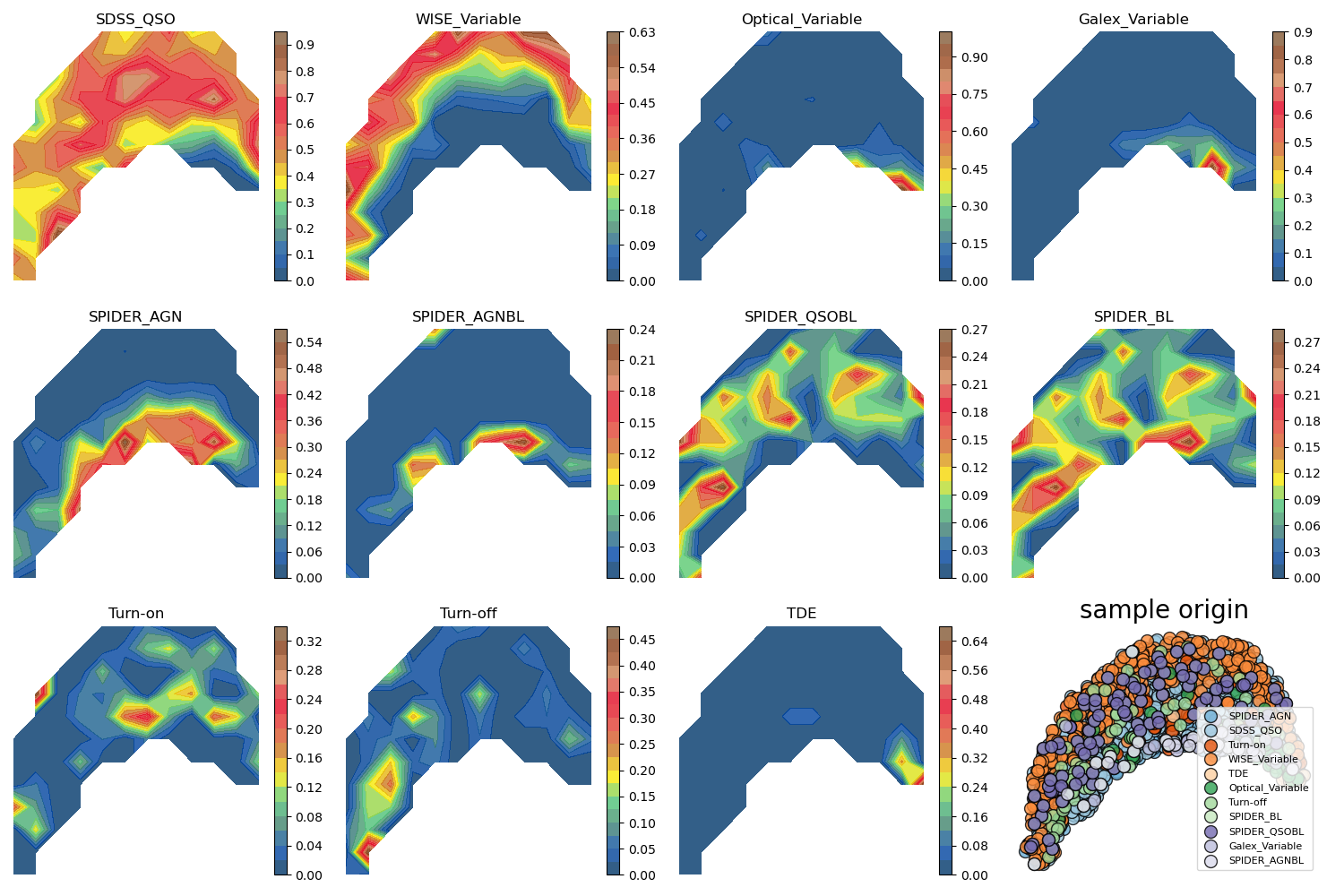}
    \caption{Separation of different labels in Sample A is illustrated using UMAPs, with parameters similar to those in Figure \ref{fig:umapztfw-main}. The bottom right panel displays the origin labels for all objects in this sample. The remaining panels feature 2D histograms, each representing the distribution of individual labels.}
    \label{fig:umapztfw-sample}
\end{figure*}

\section{Results}\label{sec:results}

Figure~\ref{fig:umapw1-main} shows the UMAP projections learned using the WISE W1 band alone. As expected given the flux normalization applied during preprocessing, no systematic trend with mean brightness is visible in the left panel. Likewise, the absence of a redshift gradient in the middle panel indicates that the manifold is primarily structured by patterns in the time-domain light curves rather than by absolute flux scale or rest-frame wavelength. Although the WISE W1 band probes rest-frame wavelengths spanning approximately 2.8--1.7~$\mu$m across Sample~A, this variation does not appear to dominate the learned variability structure. In contrast, a clear gradient in mean fractional variability is visible in the right panel, indicating that variability amplitude is one of the dominant organizing axes in this single-band embedding.

To examine how known AGN populations project onto this learned time-domain variability space, Figure~\ref{fig:umapw1-sample} shows the spatial distributions of sources with different external labels overlaid on the same UMAP embedding. The SDSS quasar subset spans much of the manifold, consistent with its heterogeneous variability behavior. In contrast, the WISE-variable AGNs preferentially populate the left-hand side of the embedding, while UV- and optically selected variables occupy overlapping but distinct regions. This pattern suggests that variability selected at different wavelengths emphasizes different aspects of the underlying time-domain behavior, although we do not assign statistical significance to these visual trends.

The middle row of Figure~\ref{fig:umapw1-sample} shows the SPIDERS subsamples. Differences between broad-line and narrow-line objects are subtle in this representation, but a qualitative tendency is visible: sources without broad emission lines preferentially occupy regions associated with lower mean fractional variability, while broad-line quasars extend into higher-variability regions. These trends are descriptive and should be interpreted cautiously, particularly given the heterogeneous selection of the SPIDERS sample.

The bottom row highlights changing-look AGNs and tidal disruption events. The Turn-On and Turn-Off CLAGNs occupy overlapping but distinct regions, particularly at higher variability amplitudes. TDEs, by contrast, cluster toward a separate edge of the manifold and do not strongly overlap with the CLAGN populations in the WISE W1-only embedding. This behavior is consistent with the extreme, transient nature of TDE variability, but the small sample size precludes firm conclusions.

The partial overlap between Turn-On CLAGNs and broad-line SPIDERS quasars is consistent with previous estimates that a non-negligible fraction of highly variable quasars may undergo changing-look transitions, depending on monitoring cadence and timescale \citep{MacLeod2016, MacLeod2019}. This overlap highlights the continuum between strong quasar variability and spectroscopically confirmed changing-look behavior.

When repeating the analysis using only a single ZTF band (the $g$ band; Figure~\ref{fig:umapzg-sample}), the distinction between Turn-On and Turn-Off CLAGNs becomes less apparent, and TDEs display more similar time-domain behavior. Given the substantially shorter observational baseline of ZTF compared to WISE, this result is consistent with the interpretation that variability timescale is an important discriminator between these populations \citep{Hon2022}. However, because most TDE searches explicitly exclude AGNs to minimize contamination, the true degree of overlap between TDEs and extreme AGN variability remains uncertain. Continued long-term monitoring of turn-on CLAGNs may help clarify this relationship.

Rather than learning the manifold from a single band, we can combine observations across facilities. Figures~\ref{fig:umapztfw-main} and \ref{fig:umapztfw-sample} show UMAP embeddings constructed from three ZTF bands together with the two WISE bands. This combination leverages the complementary strengths of the surveys: ZTF provides high-cadence sampling sensitive to short-timescale variability, while WISE provides a long temporal baseline that captures slower variability modes.

Including Pan-STARRS and Gaia data (right panel of Figure~\ref{fig:umapztfw-main}) primarily reduces the sample size, as not all objects are detected in these surveys. Because the optical wavelength coverage overlaps substantially with ZTF, this inclusion does not add significant new information to the manifold in practice, although the framework places no formal limitation on incorporating additional bands.

The combined ZTF+WISE manifold spans a broader portion of the spectral energy distribution and now exhibits a mild redshift gradient, reflecting the inclusion of multiple wavelength regimes. Variability amplitude is less dominant than in the single-band case, but still contributes to the overall structure. As shown in Figure~\ref{fig:umapztfw-sample}, UV-, optical-, and WISE-variable AGNs continue to populate partially distinct regions, though these patterns are influenced by redshift and sample composition. The SPIDERS sources again span much of the embedding, with narrow-line objects preferentially occupying lower-variability regions. The CLAGN populations are less distinctly separated in this multi-band space, likely reflecting both redshift effects and the increased dimensional complexity of the combined data. TDEs, however, remain confined to a relatively compact region at the edge of the manifold.

\begin{figure*}[ht]
    \centering
    \includegraphics[width=\textwidth]{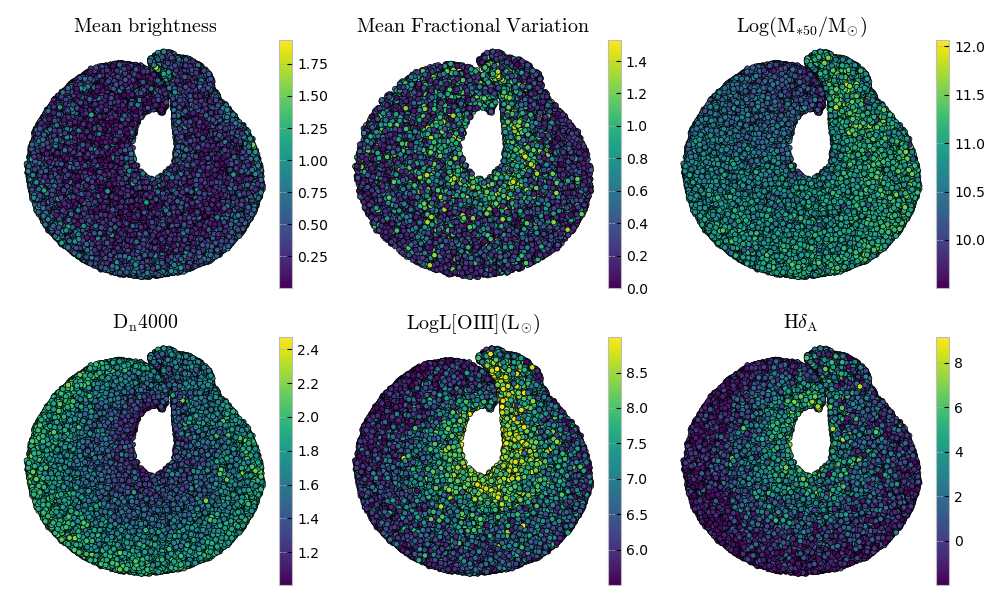}
    \caption{This figure presents the UMAPs trained on the GP regressed ZTF+WISE band lightcurves of Type-2 AGNs in Sample B. The maps from top-left to bottom-right are color coded by the mean brightness and mean fractional variation (measured from the lightcurves) and the central stellar mass($\rm Log(M_{*50}/M_{\odot})$), $\rm D_{n}4000$, $\rm [OIII]$ corrected luminosity, and $\rm H\delta_{A}$ (directly from \citealt{Kauffmann2003}).} Sample B was used here due to its uniform spectroscopic coverage, allowing us to compare the manifold structure learned from light curves with independent physical tracers of AGN activity.
    \label{fig:umapkauf}
\end{figure*}

Figure~\ref{fig:umapkauf} presents the ZTF+WISE manifold for Sample~B, the homogeneous Type-2 AGN sample drawn from \citet{Kauffmann2003}. Unlike Sample~A, this dataset provides uniform spectroscopic measurements, enabling a direct comparison between patterns in the learned time-domain space and independent physical diagnostics. Despite the narrow redshift range of this sample, the learned manifold exhibits clear gradients when colored by [O~III] luminosity, H$\delta_A$, D$_n$(4000), stellar mass, and mean fractional variability—none of which were used in constructing the embedding.

The projection reveals that [O~III] luminosity and H$\delta_A$ absorption increase toward the central regions of the manifold, while the 4000~\AA\ break weakens, indicating younger stellar populations. The mean fractional variability measured from the light curves also becomes more pronounced toward these regions. Stellar mass shows a distinct trend approximately perpendicular to these gradients, possibly reflecting differences in host-galaxy properties or dust attenuation. These orthogonal trends suggest that the manifold captures multiple, partially independent axes of structure present in the high-dimensional time-domain data.

Importantly, we do not find a strong correspondence between position on the BPT diagram and location in the variability manifold, indicating that the embedding is not simply reproducing traditional spectroscopic classifications. Instead, the manifold organizes sources according to similarities in their multi-band time-domain photometric behavior, with spectroscopic properties varying smoothly across this space. This behavior mirrors known correlations between stellar mass, stellar age, and [O~III] luminosity in Type-2 AGNs \citep{Kauffmann2003}, and demonstrates that photometric time-domain data alone encode information linked to these independently measured properties.

\section{Summary and Discussion}\label{sec:conclusions}

In this study, we explored the use of archival multi-wavelength time-series data to construct a unified representation of AGN variability using manifold learning. By integrating heterogeneous light curves from multiple surveys and applying non-linear dimensionality reduction, we show that observed time-domain variability can be organized into a low-dimensional space that captures coherent structure associated with external AGN labels and with independently measured spectroscopic properties.

The purpose of this work is to characterize similarities and differences in observed multi-band variability behavior, not to model or identify the physical mechanisms that drive AGN variability. Instead, we present a data-driven, unsupervised framework for comparing time-domain behavior across diverse samples and surveys. Using photometric light curves alone, without labels or spectroscopic inputs, we construct high-dimensional representations of AGN variability and examine their projections for two fully distinct samples: a heterogeneous compilation of variability-selected AGNs (Sample~A) and a large, homogeneous sample of Type-2 AGNs with uniform spectroscopic measurements (Sample~B).

The main results can be summarized as follows:
\begin{enumerate}
    \item Without using class labels during training, the learned manifolds organize variability-selected AGN populations into overlapping but non-identical regions. In Sample~A, WISE-variable AGNs cluster more tightly than UV- or optically selected variables, consistent with the longer variability timescales probed in the mid-infrared.

    \item Turn-on and turn-off CLAGNs occupy partially distinct regions of the manifold, indicating systematic differences in their multi-band time-domain behavior that emerge without supervision.

    \item TDEs populate restricted regions of the variability space, particularly distinct in mid-infrared bands, while showing increased overlap with CLAGNs in optical-only representations.

    \item In the homogeneous Type-2 AGN sample (Sample~B), manifold coordinates vary smoothly with independently measured spectroscopic and host-galaxy properties, including [O\,III] luminosity, D$_n$(4000), H$\delta_A$, stellar mass, and mean fractional variability, despite none of these quantities being used in constructing the manifold.
\end{enumerate}

Taken together, the results from Sample~A demonstrate that AGN populations selected using different variability criteria occupy overlapping but structured regions of variability space. Differences between mid-infrared, optical, and UV variability-selected AGNs reflect wavelength-dependent sensitivity to variability amplitude and timescale, with WISE-selected sources exhibiting more compact distributions likely driven by long-term variability. Turn-on and turn-off CLAGNs likewise show systematic offsets, while TDEs cluster in restricted regions despite not being explicitly modeled. These patterns indicate that time-domain variability alone is sufficient to recover characteristic multi-band variability signatures at a descriptive level, without requiring labels or physical inputs, though we do not assess statistical separability or define sharp population boundaries.

The clearest indication that the learned manifold captures information beyond sample labeling comes from Sample~B, which serves as an internal validation set. In this homogeneous Type-2 AGN sample, manifold coordinates correlate smoothly with spectroscopic diagnostics that trace nuclear activity and host-galaxy properties. We use the term ``physically meaningful'' in a restricted sense: the variability manifold learned from photometric time series alone captures structure that is correlated with independently measured physical quantities, without implying a causal relationship or identifying the underlying drivers of variability. This demonstrates that similarities in observed variability behavior encode information related to physical properties measured through spectroscopy.

Within the learned manifold, two broad types of structure are evident. First, sources sharing external labels often occupy similar regions, reflecting common variability behavior. Second, continuous gradients appear that correlate with spectroscopic observables, particularly in Sample~B. These structures arise solely from the time-domain data and were not imposed through feature engineering or supervised learning. At the same time, spatial proximity in the manifold reflects similarity in observed variability behavior and should not be interpreted as evidence for shared physical mechanisms or evolutionary pathways. Apparent overlap between populations often reflects similarities in variability amplitude or timescale that may arise from different underlying processes. The manifold is therefore best viewed as an exploratory diagnostic that complements, rather than replaces, traditional physical measurements.

Several limitations and directions for future work follow naturally from this analysis. First, the external labels used here are heterogeneous and not standardized across studies, particularly for CLAGNs, which are identified using diverse selection criteria and confidence thresholds. While our goal is not to validate individual classifications, this heterogeneity highlights the need for more uniform definitions in future time-domain studies.

Second, although we allow for non-Euclidean distance metrics in constructing the variability space, the use of UMAP assumes that the data lie on a Riemannian manifold. This assumption may not strictly hold for light curves, which can exhibit non-stationary behavior, multiple characteristic timescales, and more complex geometric or topological structure. Future work will explore alternative representations and quantitative diagnostics, such as neighborhood purity or completeness, to characterize structure in variability space more rigorously.

Third, the preprocessing steps required to unify asynchronous, multi-band light curves involve trade-offs. Temporal alignment simplifies analysis but may obscure physically meaningful time delays between bands. Differences between optical and infrared variability, for example, can carry important physical information that is not explicitly modeled here. Incorporating relative timing or lag-aware representations is a natural extension of this framework.

The quality and uniformity of archival photometry also remain important considerations. While manifold learning can mitigate some non-systematic uncertainties, improvements in light-curve extraction, such as consistent forced photometry, PSF-aware measurements, or denoising and resolution enhancement, would directly benefit analyses of this kind. Community-wide efforts toward standardized photometric pipelines would significantly enhance the interpretability of large time-domain datasets.

More broadly, the learned variability space provides a compact, survey-agnostic representation of time-domain behavior that can be used to compare heterogeneous samples directly. In settings where spectroscopy is unavailable or incomplete, proximity in variability space may serve as a prior for identifying sources likely to share similar physical properties. Rare populations such as CLAGNs and TDEs occupy restricted regions of the manifold, enabling the identification and prioritization of new candidates for spectroscopic follow-up based solely on their multi-band variability behavior. A quantitative assessment of targeting efficiency represents a natural next step beyond the proof-of-concept demonstration presented here.

While our analysis focuses on AGNs, the methodology is broadly applicable to other classes of variable and transient sources. Upcoming surveys such as LSST and the Roman Space Telescope will produce vast, heterogeneous time-domain datasets for which unsupervised, time-domain representations will be increasingly valuable.

Finally, we emphasize the exploratory nature of this work. We do not assess the statistical significance of the clustering patterns observed in the UMAP projections, which are used primarily for visualization and hypothesis generation. Moreover, while UMAP preserves local structure effectively, it may distort global relationships between clusters \citep{wang2021, kobak2019, kobak2021, marx2024}. Distances between widely separated regions in the embedding should therefore be interpreted with caution. Future work will focus on developing quantitative diagnostics for structure in variability space and applying this framework to larger and more diverse samples.

This work made extensive use of the Fornax Science Console, part of the NASA Astrophysics cloud-based Fornax Initiative jointly developed by Goddard Space Flight Center's Astrophysics Projects Division and the HEASARC, IRSA, and MAST archives. SH thanks R.~Chary for thoughtful discussions and feedback. We thank the referee for constructive comments that improved the clarity and scope of this work.

\textit{Software:} Astropy (\citealt{astropy2018}), ChatGPT (OpenAI, 2024-- for language editing and readability improvements), HEALPix (\citealt{Gorski2005}), scikit-learn (\citealt{scikit-learn}), UMAP (\citealt{McInnes2018})

\bibliography{AGNs.bib}

@article{keogh2005,
  title={Exact indexing of dynamic time warping},
  author={Keogh, Eamonn and Ratanamahatana, Chotirat Ann},
  journal={Knowledge and information systems},
  volume={7},
  number={3},
  pages={358--386},
  year={2005},
  publisher={Springer}
}

@ARTICLE{Sanjaripour2025,
       author = {{Sanjaripour}, Sogol and {Aravindan}, Archana and {Canalizo}, Gabriela and {Hemmati}, Shoubaneh and {Mobasher}, Bahram and {Coil}, Alison L. and {Barish}, Barry C.},
        title = "{Selection of Dwarf Galaxies Hosting Active Galactic Nuclei: A Measure of Bias and Contamination Using Unsupervised Machine Learning Techniques}",
      journal = {\apj},
         year = 2025,
        month = oct,
       volume = {992},
       number = {1},
          eid = {138},
        pages = {138},
          doi = {10.3847/1538-4357/ae0326},
archivePrefix = {arXiv},
       eprint = {2505.16509},
 primaryClass = {astro-ph.GA},
       adsurl = {https://ui.adsabs.harvard.edu/abs/2025ApJ...992..138S}
}

@article{marx2024,
  title={Seeing data as t-SNE and UMAP do},
  author={Marx, Vivien},
  journal={Nature Methods},
  volume={21},
  number={6},
  pages={930--933},
  year={2024},
  publisher={Nature Publishing Group US New York}
}

@article{kobak2019,
  title={The art of using t-SNE for single-cell transcriptomics},
  author={Kobak, Dmitry and Berens, Philipp},
  journal={Nature communications},
  volume={10},
  number={1},
  pages={5416},
  year={2019},
  publisher={Nature Publishing Group UK London}
}

@article{kobak2021,
  title={Initialization is critical for preserving global data structure in both t-SNE and UMAP},
  author={Kobak, Dmitry and Linderman, George C},
  journal={Nature biotechnology},
  volume={39},
  number={2},
  pages={156--157},
  year={2021},
  publisher={Nature Publishing Group US New York}
}

@misc{wang2021,
      title={Understanding How Dimension Reduction Tools Work: An Empirical Approach to Deciphering t-SNE, UMAP, TriMAP, and PaCMAP for Data Visualization}, 
      author={Yingfan Wang and Haiyang Huang and Cynthia Rudin and Yaron Shaposhnik},
      year={2021},
      eprint={2012.04456},
      archivePrefix={arXiv},
      primaryClass={cs.LG},
      url={https://arxiv.org/abs/2012.04456}, 
}

@article{Beck2018,
   title={R Package imputeTestbench to Compare Imputation Methods for Univariate Time Series},
   volume={10},
   ISSN={2073-4859},
   url={http://dx.doi.org/10.32614/RJ-2018-024},
   DOI={10.32614/rj-2018-024},
   number={1},
   journal={The R Journal},
   publisher={The R Foundation},
   author={Beck, Marcus, W and Bokde, Neeraj and Asencio-Cortés, Gualberto and Kulat, Kishore},
   year={2018},
   pages={218} }

@misc{moritz2015,
      title={Comparison of different Methods for Univariate Time Series Imputation in R}, 
      author={Steffen Moritz and Alexis Sardá and Thomas Bartz-Beielstein and Martin Zaefferer and Jörg Stork},
      year={2015},
      eprint={1510.03924},
      archivePrefix={arXiv},
      primaryClass={stat.AP},
      url={https://arxiv.org/abs/1510.03924}, 
}

@ARTICLE{Elorrieta2021,
       author = {{Elorrieta}, Felipe and {Eyheramendy}, Susana and {Palma}, Wilfredo and {Ojeda}, Cesar},
        title = "{A novel bivariate autoregressive model for predicting and forecasting irregularly observed time series}",
      journal = {\mnras},
         year = 2021,
        month = jul,
       volume = {505},
       number = {1},
        pages = {1105-1116},
          doi = {10.1093/mnras/stab1216},
archivePrefix = {arXiv},
       eprint = {2104.12248},
 primaryClass = {astro-ph.IM},
       adsurl = {https://ui.adsabs.harvard.edu/abs/2021MNRAS.505.1105E}
}

@article{Hu2020,
doi = {10.3847/1538-3881/abc1e2},
url = {https://doi.org/10.3847/1538-3881/abc1e2},
year = {2020},
month = {nov},
publisher = {The American Astronomical Society},
volume = {160},
number = {6},
pages = {265},
author = {Hu, Zhirui and Tak, Hyungsuk},
title = {Modeling Stochastic Variability in Multiband Time-series Data},
journal = {The Astronomical Journal}
}

@ARTICLE{Sanjaripour2024,
       author = {{Sanjaripour}, Sogol and {Hemmati}, Shoubaneh and {Mobasher}, Bahram and {Canalizo}, Gabriela and {Barish}, Barry C. and {Shivaei}, Irene and {Coil}, Alison L. and {Chartab}, Nima and {Jafariyazani}, Marziye and {Reddy}, Naveen A. and {Azadi}, Mojegan},
        title = "{The Application of Manifold Learning to a Selection of Different Galaxy Populations and Scaling Relation Analysis}",
      journal = {\apj},
         year = 2024,
        month = dec,
       volume = {977},
       number = {2},
          eid = {202},
        pages = {202},
          doi = {10.3847/1538-4357/ad90ba},
archivePrefix = {arXiv},
       eprint = {2410.07354},
 primaryClass = {astro-ph.GA},
       adsurl = {https://ui.adsabs.harvard.edu/abs/2024ApJ...977..202S}
}

@ARTICLE{Wang2025,
       author = {{Wang}, Shu and {Woo}, Jong-Hak and {Gallo}, Elena and {Son}, Donghoon and {Yang}, Qian and {Jin}, Junjie and {Guo}, Hengxiao and {Kong}, Minzhi},
        title = "{Dormancy and Reawakening over Years: Eight New Recurrent Changing-look AGNs}",
      journal = {\apj},
         year = 2025,
        month = mar,
       volume = {981},
       number = {2},
          eid = {129},
        pages = {129},
          doi = {10.3847/1538-4357/adadf3},
archivePrefix = {arXiv},
       eprint = {2410.15587},
 primaryClass = {astro-ph.GA},
       adsurl = {https://ui.adsabs.harvard.edu/abs/2025ApJ...981..129W}
}

@ARTICLE{Yang2025,
       author = {{Yang}, Qian and {Green}, Paul J. and {Wu}, Xue-Bing and {Eracleous}, Michael and {Jiang}, Linhua and {Fu}, Yuming},
        title = "{Galaxies Lighting Up: Discovery of Seventy New Turn-on Changing-look Active Galactic Nuclei}",
      journal = {\apj},
         year = 2025,
        month = feb,
       volume = {980},
       number = {1},
          eid = {91},
        pages = {91},
          doi = {10.3847/1538-4357/ad94ed},
archivePrefix = {arXiv},
       eprint = {2408.16183},
 primaryClass = {astro-ph.GA},
       adsurl = {https://ui.adsabs.harvard.edu/abs/2025ApJ...980...91Y}
}

@ARTICLE{Hon2022,
       author = {{Hon}, Wei Jeat and {Wolf}, Christian and {Onken}, Christopher A. and {Webster}, Rachel and {Auchettl}, Katie},
        title = "{SkyMapper colours of Seyfert galaxies and changing-look AGN - II. Newly discovered changing-look AGN}",
      journal = {\mnras},
         year = 2022,
        month = mar,
       volume = {511},
       number = {1},
        pages = {54-70},
          doi = {10.1093/mnras/stab3694},
       adsurl = {https://ui.adsabs.harvard.edu/abs/2022MNRAS.511...54H}
}

@ARTICLE{Wang2024,
       author = {{Wang}, Yihan and {Graham}, Matthew J. and {Ford}, K.~E. Saavik and {McKernan}, Barry and {Ryu}, Taeho and {Stern}, Daniel},
        title = "{Conditions for Changing-Look AGNs from Accretion Disk-Induced Tidal Disruption Events}",
      journal = {arXiv e-prints},
         year = 2024,
        month = jun,
          eid = {arXiv:2406.12096},
        pages = {arXiv:2406.12096},
          doi = {10.48550/arXiv.2406.12096},
archivePrefix = {arXiv},
       eprint = {2406.12096},
 primaryClass = {astro-ph.HE},
       adsurl = {https://ui.adsabs.harvard.edu/abs/2024arXiv240612096W}
}

@ARTICLE{Yang2023,
       author = {{Yang}, Qian and {Green}, Paul J. and {MacLeod}, Chelsea L. and {Plotkin}, Richard M. and {Anderson}, Scott F. and {Bieryla}, Allyson and {Civano}, Francesca and {Eracleous}, Michael and {Graham}, Matthew and {Ruan}, John J. and {Runnoe}, Jessie and {Zhao}, Xiurui},
        title = "{Probing the Origin of Changing-look Quasar Transitions with Chandra}",
      journal = {\apj},
         year = 2023,
        month = aug,
       volume = {953},
       number = {1},
          eid = {61},
        pages = {61},
          doi = {10.3847/1538-4357/acdedd},
archivePrefix = {arXiv},
       eprint = {2303.06733},
 primaryClass = {astro-ph.GA},
       adsurl = {https://ui.adsabs.harvard.edu/abs/2023ApJ...953...61Y}
}

@ARTICLE{Stern2018,
       author = {{Stern}, Daniel and {McKernan}, Barry and {Graham}, Matthew J. and {Ford}, K.~E.~S. and {Ross}, Nicholas P. and {Meisner}, Aaron M. and {Assef}, Roberto J. and {Balokovi{\'c}}, Mislav and {Brightman}, Murray and {Dey}, Arjun and {Drake}, Andrew and {Djorgovski}, S.~G. and {Eisenhardt}, Peter and {Jun}, Hyunsung D.},
        title = "{A Mid-IR Selected Changing-look Quasar and Physical Scenarios for Abrupt AGN Fading}",
      journal = {\apj},
         year = 2018,
        month = sep,
       volume = {864},
       number = {1},
          eid = {27},
        pages = {27},
          doi = {10.3847/1538-4357/aac726},
archivePrefix = {arXiv},
       eprint = {1805.06920},
 primaryClass = {astro-ph.GA},
       adsurl = {https://ui.adsabs.harvard.edu/abs/2018ApJ...864...27S}
}

@ARTICLE{Chan2019,
       author = {{Chan}, Chi-Ho and {Piran}, Tsvi and {Krolik}, Julian H. and {Saban}, Dekel},
        title = "{Tidal Disruption Events in Active Galactic Nuclei}",
      journal = {\apj},
         year = 2019,
        month = aug,
       volume = {881},
       number = {2},
          eid = {113},
        pages = {113},
          doi = {10.3847/1538-4357/ab2b40},
archivePrefix = {arXiv},
       eprint = {1904.12261},
 primaryClass = {astro-ph.HE},
       adsurl = {https://ui.adsabs.harvard.edu/abs/2019ApJ...881..113C}
}

@ARTICLE{Ricci2023,
       author = {{Ricci}, Claudio and {Trakhtenbrot}, Benny},
        title = "{Changing-look active galactic nuclei}",
      journal = {Nature Astronomy},
         year = 2023,
        month = nov,
       volume = {7},
        pages = {1282-1294},
          doi = {10.1038/s41550-023-02108-4},
archivePrefix = {arXiv},
       eprint = {2211.05132},
 primaryClass = {astro-ph.GA},
       adsurl = {https://ui.adsabs.harvard.edu/abs/2023NatAs...7.1282R}
}

@ARTICLE{Ricci2020,
       author = {{Ricci}, C. and {Kara}, E. and {Loewenstein}, M. and {Trakhtenbrot}, B. and {Arcavi}, I. and {Remillard}, R. and {Fabian}, A.~C. and {Gendreau}, K.~C. and {Arzoumanian}, Z. and {Li}, R. and {Ho}, L.~C. and {MacLeod}, C.~L. and {Cackett}, E. and {Altamirano}, D. and {Gandhi}, P. and {Kosec}, P. and {Pasham}, D. and {Steiner}, J. and {Chan}, C. -H.},
        title = "{The Destruction and Recreation of the X-Ray Corona in a Changing-look Active Galactic Nucleus}",
      journal = {\apjl},
         year = 2020,
        month = jul,
       volume = {898},
       number = {1},
          eid = {L1},
        pages = {L1},
          doi = {10.3847/2041-8213/ab91a1},
archivePrefix = {arXiv},
       eprint = {2007.07275},
 primaryClass = {astro-ph.HE},
       adsurl = {https://ui.adsabs.harvard.edu/abs/2020ApJ...898L...1R}
}

@ARTICLE{Neustadt2020,
       author = {{Neustadt}, J.~M.~M. and {Holoien}, T.~W. -S. and {Kochanek}, C.~S. and {Auchettl}, K. and {Brown}, J.~S. and {Shappee}, B.~J. and {Pogge}, R.~W. and {Dong}, Subo and {Stanek}, K.~Z. and {Tucker}, M.~A. and {Bose}, S. and {Chen}, Ping and {Ricci}, C. and {Vallely}, P.~J. and {Prieto}, J.~L. and {Thompson}, T.~A. and {Coulter}, D.~A. and {Drout}, M.~R. and {Foley}, R.~J. and {Kilpatrick}, C.~D. and {Piro}, A.~L. and {Rojas-Bravo}, C. and {Buckley}, D.~A.~H. and {Gromadzki}, M. and {Dimitriadis}, G. and {Siebert}, M.~R. and {Do}, A. and {Huber}, M.~E. and {Payne}, A.~V.},
        title = "{To TDE or not to TDE: the luminous transient ASASSN-18jd with TDE-like and AGN-like qualities}",
      journal = {\mnras},
         year = 2020,
        month = may,
       volume = {494},
       number = {2},
        pages = {2538-2560},
          doi = {10.1093/mnras/staa859},
archivePrefix = {arXiv},
       eprint = {1910.01142},
 primaryClass = {astro-ph.HE},
       adsurl = {https://ui.adsabs.harvard.edu/abs/2020MNRAS.494.2538N}
}

@article{Zakamska2003,
  author = {Zakamska, N. L. and Strauss, M. A. and Krolik, J. H. and et al.},
  title = {Type II Quasars from the Sloan Digital Sky Survey: I. Sample Selection},
  journal = {AJ},
  year = {2003},
  volume = {126},
  pages = {2125--2144},
  doi = {10.1086/378610}
}

@article{Reyes2008,
  author = {Reyes, R. and Zakamska, N. L. and Strauss, M. A. and et al.},
  title = {Space Density of Optically Selected Type 2 Quasars},
  journal = {AJ},
  year = {2008},
  volume = {136},
  pages = {2373--2390},
  doi = {10.1088/0004-6256/136/6/2373}
}

@article{Ross2020,
  author = {Ross, N. P. and Ford, K. E. S. and Graham, M. J. and Stern, D. and Drake, A. J. and Djorgovski, S. G. and Mahabal, A. and Nugent, P. and Palaversa, L. and Ruan, J. J. and others},
  title = {A Systematic Search for Changing-look Quasars in SDSS and PS1},
  journal = {Astrophysical Journal},
  volume = {903},
  number = {2},
  pages = {150},
  year = {2020},
  doi = {10.3847/1538-4357/abb9b6}
}

@article{MacLeod2010,
  author = {MacLeod, C. L. and Ivezić, Z. and Kochanek, C. S. and Kozłowski, S. and Kelly, B. C. and Bullock, E. and Kimball, A. and Sesar, B. and Westman, D. and Brooks, K. and Gibson, R. and Becker, A. C. and de Vries, W. H.},
  title = {Modeling the Time Variability of SDSS Stripe 82 Quasars as a Damped Random Walk},
  journal = {The Astrophysical Journal},
  volume = {721},
  number = {2},
  pages = {1014--1033},
  year = {2010},
  doi = {10.1088/0004-637X/721/2/1014}
}

@article{McHardy2006,
  author = {McHardy, I. M. et al.},
  title = {Active galactic nuclei as scaled-up Galactic black holes},
  journal = {Nature},
  volume = {444},
  pages = {730--732},
  year = {2006},
  doi = {10.1038/nature05389}
}

@article{Koshida2014,
  author = {Koshida, S. et al.},
  title = {Reverberation Measurements of the Inner Radius of the Dust Torus in 17 Seyfert Galaxies},
  journal = {Astrophysical Journal},
  volume = {788},
  number = {2},
  pages = {159},
  year = {2014},
  doi = {10.1088/0004-637X/788/2/159}
}

@ARTICLE{Antonucci1984,
       author = {{Antonucci}, R.~R.~J.},
        title = "{Optical spectropolarimetry of radio galaxies.}",
      journal = {\apj},
         year = 1984,
        month = mar,
       volume = {278},
        pages = {499-520},
          doi = {10.1086/161816},
       adsurl = {https://ui.adsabs.harvard.edu/abs/1984ApJ...278..499A}
}

@ARTICLE{Masters2015,
       author = {{Masters}, Daniel and {Capak}, Peter and {Stern}, Daniel and {Ilbert}, Olivier and {Salvato}, Mara and {Schmidt}, Samuel and {Longo}, Giuseppe and {Rhodes}, Jason and {Paltani}, Stephane and {Mobasher}, Bahram and {Hoekstra}, Henk and {Hildebrandt}, Hendrik and {Coupon}, Jean and {Steinhardt}, Charles and {Speagle}, Josh and {Faisst}, Andreas and {Kalinich}, Adam and {Brodwin}, Mark and {Brescia}, Massimo and {Cavuoti}, Stefano},
        title = "{Mapping the Galaxy Color-Redshift Relation: Optimal Photometric Redshift Calibration Strategies for Cosmology Surveys}",
      journal = {\apj},
         year = 2015,
        month = nov,
       volume = {813},
       number = {1},
          eid = {53},
        pages = {53},
          doi = {10.1088/0004-637X/813/1/53},
archivePrefix = {arXiv},
       eprint = {1509.03318},
 primaryClass = {astro-ph.CO},
       adsurl = {https://ui.adsabs.harvard.edu/abs/2015ApJ...813...53M}
}

@ARTICLE{Kelly2009,
       author = {{Kelly}, Brandon C. and {Bechtold}, Jill and {Siemiginowska}, Aneta},
        title = "{Are the Variations in Quasar Optical Flux Driven by Thermal Fluctuations?}",
      journal = {\apj},
         year = 2009,
        month = jun,
       volume = {698},
       number = {1},
        pages = {895-910},
          doi = {10.1088/0004-637X/698/1/895},
archivePrefix = {arXiv},
       eprint = {0903.5315},
 primaryClass = {astro-ph.CO},
       adsurl = {https://ui.adsabs.harvard.edu/abs/2009ApJ...698..895K}
}

@ARTICLE{Burke2021,
       author = {{Burke}, Colin J. and {Shen}, Yue and {Blaes}, Omer and {Gammie}, Charles F. and {Horne}, Keith and {Jiang}, Yan-Fei and {Liu}, Xin and {McHardy}, Ian M. and {Morgan}, Christopher W. and {Scaringi}, Simone and {Yang}, Qian},
        title = "{A characteristic optical variability time scale in astrophysical accretion disks}",
      journal = {Science},
         year = 2021,
        month = aug,
       volume = {373},
       number = {6556},
        pages = {789-792},
          doi = {10.1126/science.abg9933},
archivePrefix = {arXiv},
       eprint = {2108.05389},
 primaryClass = {astro-ph.GA},
       adsurl = {https://ui.adsabs.harvard.edu/abs/2021Sci...373..789B}
}

@ARTICLE{Stern2005,
       author = {{Stern}, Daniel and {Eisenhardt}, Peter and {Gorjian}, Varoujan and {Kochanek}, Christopher S. and {Caldwell}, Nelson and {Eisenstein}, Daniel and {Brodwin}, Mark and {Brown}, Michael J.~I. and {Cool}, Richard and {Dey}, Arjun and {Green}, Paul and {Jannuzi}, Buell T. and {Murray}, Stephen S. and {Pahre}, Michael A. and {Willner}, S.~P.},
        title = "{Mid-Infrared Selection of Active Galaxies}",
      journal = {\apj},
         year = 2005,
        month = sep,
       volume = {631},
       number = {1},
        pages = {163-168},
          doi = {10.1086/432523},
archivePrefix = {arXiv},
       eprint = {astro-ph/0410523},
 primaryClass = {astro-ph},
       adsurl = {https://ui.adsabs.harvard.edu/abs/2005ApJ...631..163S}
}

@article{Antonucci1993,
    author = {Antonucci, R.},
    title = {Unified models for active galactic nuclei and quasars},
    journal = {Annual Review of Astronomy and Astrophysics},
    volume = {31},
    year = {1993},
    pages = {473–521},
    doi = {10.1146/annurev.aa.31.090193.002353}
}

@article{Urry1995,
    author = {Urry, C. M. and Padovani, P.},
    title = {Unified Schemes for Radio-Loud Active Galactic Nuclei},
    journal = {Publications of the Astronomical Society of the Pacific},
    volume = {107},
    year = {1995},
    pages = {803–845},
    doi = {10.1086/133630}
}

@book{Netzer2015,
    author = {Netzer, H.},
    title = {The Physics and Evolution of Active Galactic Nuclei},
    publisher = {Cambridge University Press},
    year = {2015},
    isbn = {978-1-107-03432-1}
}

@article{Ulrich1997,
    author = {Ulrich, M.-H. and Maraschi, L. and Urry, C. M.},
    title = {Variability of Active Galactic Nuclei},
    journal = {Annual Review of Astronomy and Astrophysics},
    volume = {35},
    year = {1997},
    pages = {445–502},
    doi = {10.1146/annurev.astro.35.1.445}
}

@article{Bentz2013,
    author = {Bentz, M. C. and Denney, K. D. and Grier, C. J. and Barth, A. J. and Peterson, B. M. and Vestergaard, M. and Bennert, V. N.},
    title = {The Low-luminosity End of the Radius-Luminosity Relationship for Active Galactic Nuclei},
    journal = {Astrophysical Journal},
    volume = {767},
    year = {2013},
    pages = {149},
    doi = {10.1088/0004-637X/767/2/149}
}

@ARTICLE{Sokolovsky2017,
       author = {{Sokolovsky}, K.~V. and {Gavras}, P. and {Karampelas}, A. and {Antipin}, S.~V. and {Bellas-Velidis}, I. and {Benni}, P. and {Bonanos}, A.~Z. and {Burdanov}, A.~Y. and {Derlopa}, S. and {Hatzidimitriou}, D. and {Khokhryakova}, A.~D. and {Kolesnikova}, D.~M. and {Korotkiy}, S.~A. and {Lapukhin}, E.~G. and {Moretti}, M.~I. and {Popov}, A.~A. and {Pouliasis}, E. and {Samus}, N.~N. and {Spetsieri}, Z. and {Veselkov}, S.~A. and {Volkov}, K.~V. and {Yang}, M. and {Zubareva}, A.~M.},
        title = "{Comparative performance of selected variability detection techniques in photometric time series data}",
      journal = {\mnras},
         year = 2017,
        month = jan,
       volume = {464},
       number = {1},
        pages = {274-292},
          doi = {10.1093/mnras/stw2262},
archivePrefix = {arXiv},
       eprint = {1609.01716},
 primaryClass = {astro-ph.IM},
       adsurl = {https://ui.adsabs.harvard.edu/abs/2017MNRAS.464..274S}
}

@ARTICLE{Pantoja2022,
       author = {{Pantoja}, R. and {Catelan}, M. and {Pichara}, K. and {Protopapas}, P.},
        title = "{Semi-supervised classification and clustering analysis for variable stars}",
      journal = {\mnras},
         year = 2022,
        month = dec,
       volume = {517},
       number = {3},
        pages = {3660-3681},
          doi = {10.1093/mnras/stac2715},
archivePrefix = {arXiv},
       eprint = {2209.09957},
 primaryClass = {astro-ph.SR},
       adsurl = {https://ui.adsabs.harvard.edu/abs/2022MNRAS.517.3660P}
}

@ARTICLE{Faisst2019,
       author = {{Faisst}, Andreas L. and {Prakash}, Abhishek and {Capak}, Peter L. and {Lee}, Bomee},
        title = "{How to Find Variable Active Galactic Nuclei with Machine Learning}",
      journal = {\apjl},
         year = 2019,
        month = aug,
       volume = {881},
       number = {1},
          eid = {L9},
        pages = {L9},
          doi = {10.3847/2041-8213/ab3581},
archivePrefix = {arXiv},
       eprint = {1908.07542},
 primaryClass = {astro-ph.IM},
       adsurl = {https://ui.adsabs.harvard.edu/abs/2019ApJ...881L...9F}
}

@ARTICLE{Cackett2021,
       author = {{Cackett}, Edward M. and {Bentz}, Misty C. and {Kara}, Erin},
        title = "{Reverberation mapping of active galactic nuclei: from X-ray corona to dusty torus}",
      journal = {iScience},
         year = 2021,
        month = jun,
       volume = {24},
       number = {6},
        pages = {102557},
          doi = {10.1016/j.isci.2021.102557},
archivePrefix = {arXiv},
       eprint = {2105.06926},
 primaryClass = {astro-ph.GA},
       adsurl = {https://ui.adsabs.harvard.edu/abs/2021iSci...24j2557C}
}

@ARTICLE{Shappee2014,
       author = {{Shappee}, B.~J. and {Prieto}, J.~L. and {Grupe}, D. and {Kochanek}, C.~S. and {Stanek}, K.~Z. and {De Rosa}, G. and {Mathur}, S. and {Zu}, Y. and {Peterson}, B.~M. and {Pogge}, R.~W. and {Komossa}, S. and {Im}, M. and {Jencson}, J. and {Holoien}, T.~W. -S. and {Basu}, U. and {Beacom}, J.~F. and {Szczygie{\l}}, D.~M. and {Brimacombe}, J. and {Adams}, S. and {Campillay}, A. and {Choi}, C. and {Contreras}, C. and {Dietrich}, M. and {Dubberley}, M. and {Elphick}, M. and {Foale}, S. and {Giustini}, M. and {Gonzalez}, C. and {Hawkins}, E. and {Howell}, D.~A. and {Hsiao}, E.~Y. and {Koss}, M. and {Leighly}, K.~M. and {Morrell}, N. and {Mudd}, D. and {Mullins}, D. and {Nugent}, J.~M. and {Parrent}, J. and {Phillips}, M.~M. and {Pojmanski}, G. and {Rosing}, W. and {Ross}, R. and {Sand}, D. and {Terndrup}, D.~M. and {Valenti}, S. and {Walker}, Z. and {Yoon}, Y.},
        title = "{The Man behind the Curtain: X-Rays Drive the UV through NIR Variability in the 2013 Active Galactic Nucleus Outburst in NGC 2617}",
      journal = {\apj},
         year = 2014,
        month = jun,
       volume = {788},
       number = {1},
          eid = {48},
        pages = {48},
          doi = {10.1088/0004-637X/788/1/48},
archivePrefix = {arXiv},
       eprint = {1310.2241},
 primaryClass = {astro-ph.HE},
       adsurl = {https://ui.adsabs.harvard.edu/abs/2014ApJ...788...48S}
}

@ARTICLE{Prakash2019,
       author = {{Prakash}, Abhishek and {Chary}, Ranga Ram and {Helou}, George and {Faisst}, Andreas and {Graham}, Matthew J. and {Masci}, Frank J. and {Shupe}, David L. and {Lee}, Bomee},
        title = "{A Flaring AGN in a ULIRG Candidate in Stripe 82}",
      journal = {\apj},
         year = 2019,
        month = oct,
       volume = {883},
       number = {2},
          eid = {154},
        pages = {154},
          doi = {10.3847/1538-4357/ab3b0b},
archivePrefix = {arXiv},
       eprint = {1908.04280},
 primaryClass = {astro-ph.GA},
       adsurl = {https://ui.adsabs.harvard.edu/abs/2019ApJ...883..154P}
}

@INPROCEEDINGS{Peterson2001,
       author = {{Peterson}, Bradley M.},
        title = "{Variability of Active Galactic Nuclei}",
    booktitle = {Advanced Lectures on the Starburst-AGN},
         year = 2001,
       editor = {{Aretxaga}, Itziar and {Kunth}, Daniel and {M{\'u}jica}, Ra{\'u}l},
        month = jan,
        pages = {3},
          doi = {10.1142/9789812811318_0002},
archivePrefix = {arXiv},
       eprint = {astro-ph/0109495},
 primaryClass = {astro-ph},
       adsurl = {https://ui.adsabs.harvard.edu/abs/2001sac..conf....3P}
}

@ARTICLE{Hemmati2019a,
       author = {{Hemmati}, Shoubaneh and {Capak}, Peter and {Masters}, Daniel and {Davidzon}, Iary and {Dor{\`e}}, Olivier and {Kruk}, Jeffrey and {Mobasher}, Bahram and {Rhodes}, Jason and {Scolnic}, Daniel and {Stern}, Daniel},
        title = "{Photometric Redshift Calibration Requirements for WFIRST Weak-lensing Cosmology: Predictions from CANDELS}",
      journal = {\apj},
         year = 2019,
        month = jun,
       volume = {877},
       number = {2},
          eid = {117},
        pages = {117},
          doi = {10.3847/1538-4357/ab1be5},
archivePrefix = {arXiv},
       eprint = {1808.10458},
 primaryClass = {astro-ph.GA},
       adsurl = {https://ui.adsabs.harvard.edu/abs/2019ApJ...877..117H}
}

@ARTICLE{Hemmati2019b,
       author = {{Hemmati}, Shoubaneh and {Capak}, Peter and {Pourrahmani}, Milad and {Nayyeri}, Hooshang and {Stern}, Daniel and {Mobasher}, Bahram and {Darvish}, Behnam and {Davidzon}, Iary and {Ilbert}, Olivier and {Masters}, Daniel and {Shahidi}, Abtin},
        title = "{Bringing Manifold Learning and Dimensionality Reduction to SED Fitters}",
      journal = {\apjl},
         year = 2019,
        month = aug,
       volume = {881},
       number = {1},
          eid = {L14},
        pages = {L14},
          doi = {10.3847/2041-8213/ab3418},
archivePrefix = {arXiv},
       eprint = {1905.10379},
 primaryClass = {astro-ph.GA},
       adsurl = {https://ui.adsabs.harvard.edu/abs/2019ApJ...881L..14H}
}

@ARTICLE{BPT1981,
       author = {{Baldwin}, J.~A. and {Phillips}, M.~M. and {Terlevich}, R.},
        title = "{Classification parameters for the emission-line spectra of extragalactic objects.}",
      journal = {\pasp},
         year = 1981,
        month = feb,
       volume = {93},
        pages = {5-19},
          doi = {10.1086/130766},
       adsurl = {https://ui.adsabs.harvard.edu/abs/1981PASP...93....5B}
}

@ARTICLE{Somalwar2023,
       author = {{Somalwar}, Jean J. and {Ravi}, Vikram and {Dong}, Dillon Z. and {Chen}, Yuyang and {Breen}, Shari and {Chandra}, Poonam and {Clarke}, Tracy and {De}, Kishalay and {Gaensler}, B.~M. and {Hallinan}, Gregg and {Laha}, Sibasish and {Law}, Casey and {Myers}, Steven T. and {Parsotan}, Tyler and {Peters}, Wendy and {Polisensky}, Emil},
        title = "{A Candidate Relativistic Tidal Disruption Event at 340 Mpc}",
      journal = {\apj},
         year = 2023,
        month = mar,
       volume = {945},
       number = {2},
          eid = {142},
        pages = {142},
          doi = {10.3847/1538-4357/acbafc},
archivePrefix = {arXiv},
       eprint = {2207.02873},
 primaryClass = {astro-ph.HE},
       adsurl = {https://ui.adsabs.harvard.edu/abs/2023ApJ...945..142S}
}

@ARTICLE{Hammerstein2023,
       author = {{Hammerstein}, Erica and {van Velzen}, Sjoert and {Gezari}, Suvi and {Cenko}, S. Bradley and {Yao}, Yuhan and {Ward}, Charlotte and {Frederick}, Sara and {Villanueva}, Natalia and {Somalwar}, Jean J. and {Graham}, Matthew J. and {Kulkarni}, Shrinivas R. and {Stern}, Daniel and {Andreoni}, Igor and {Bellm}, Eric C. and {Dekany}, Richard and {Dhawan}, Suhail and {Drake}, Andrew J. and {Fremling}, Christoffer and {Gatkine}, Pradip and {Groom}, Steven L. and {Ho}, Anna Y.~Q. and {Kasliwal}, Mansi M. and {Karambelkar}, Viraj and {Kool}, Erik C. and {Masci}, Frank J. and {Medford}, Michael S. and {Perley}, Daniel A. and {Purdum}, Josiah and {van Roestel}, Jan and {Sharma}, Yashvi and {Sollerman}, Jesper and {Taggart}, Kirsty and {Yan}, Lin},
        title = "{The Final Season Reimagined: 30 Tidal Disruption Events from the ZTF-I Survey}",
      journal = {\apj},
         year = 2023,
        month = jan,
       volume = {942},
       number = {1},
          eid = {9},
        pages = {9},
          doi = {10.3847/1538-4357/aca283},
archivePrefix = {arXiv},
       eprint = {2203.01461},
 primaryClass = {astro-ph.HE},
       adsurl = {https://ui.adsabs.harvard.edu/abs/2023ApJ...942....9H}
}

@ARTICLE{vanVelzen2021,
       author = {{van Velzen}, Sjoert and {Gezari}, Suvi and {Hammerstein}, Erica and {Roth}, Nathaniel and {Frederick}, Sara and {Ward}, Charlotte and {Hung}, Tiara and {Cenko}, S. Bradley and {Stein}, Robert and {Perley}, Daniel A. and {Taggart}, Kirsty and {Foley}, Ryan J. and {Sollerman}, Jesper and {Blagorodnova}, Nadejda and {Andreoni}, Igor and {Bellm}, Eric C. and {Brinnel}, Valery and {De}, Kishalay and {Dekany}, Richard and {Feeney}, Michael and {Fremling}, Christoffer and {Giomi}, Matteo and {Golkhou}, V. Zach and {Graham}, Matthew J. and {Ho}, Anna. Y.~Q. and {Kasliwal}, Mansi M. and {Kilpatrick}, Charles D. and {Kulkarni}, Shrinivas R. and {Kupfer}, Thomas and {Laher}, Russ R. and {Mahabal}, Ashish and {Masci}, Frank J. and {Miller}, Adam A. and {Nordin}, Jakob and {Riddle}, Reed and {Rusholme}, Ben and {van Santen}, Jakob and {Sharma}, Yashvi and {Shupe}, David L. and {Soumagnac}, Maayane T.},
        title = "{Seventeen Tidal Disruption Events from the First Half of ZTF Survey Observations: Entering a New Era of Population Studies}",
      journal = {\apj},
         year = 2021,
        month = feb,
       volume = {908},
       number = {1},
          eid = {4},
        pages = {4},
          doi = {10.3847/1538-4357/abc258},
archivePrefix = {arXiv},
       eprint = {2001.01409},
 primaryClass = {astro-ph.HE},
       adsurl = {https://ui.adsabs.harvard.edu/abs/2021ApJ...908....4V}
}

@ARTICLE{Clerc2016,
       author = {{Clerc}, N. and {Merloni}, A. and {Zhang}, Y. -Y. and {Finoguenov}, A. and {Dwelly}, T. and {Nandra}, K. and {Collins}, C. and {Dawson}, K. and {Kneib}, J. -P. and {Rozo}, E. and {Rykoff}, E. and {Sadibekova}, T. and {Brownstein}, J. and {Lin}, Y. -T. and {Ridl}, J. and {Salvato}, M. and {Schwope}, A. and {Steinmetz}, M. and {Seo}, H. -J. and {Tinker}, J.},
        title = "{SPIDERS: the spectroscopic follow-up of X-ray selected clusters of galaxies in SDSS-IV}",
      journal = {\mnras},
         year = 2016,
        month = dec,
       volume = {463},
       number = {4},
        pages = {4490-4515},
          doi = {10.1093/mnras/stw2214},
archivePrefix = {arXiv},
       eprint = {1608.08963},
 primaryClass = {astro-ph.CO},
       adsurl = {https://ui.adsabs.harvard.edu/abs/2016MNRAS.463.4490C}
}

@ARTICLE{Salvato2018,
       author = {{Salvato}, M. and {Buchner}, J. and {Budav{\'a}ri}, T. and {Dwelly}, T. and {Merloni}, A. and {Brusa}, M. and {Rau}, A. and {Fotopoulou}, S. and {Nandra}, K.},
        title = "{Finding counterparts for all-sky X-ray surveys with NWAY: a Bayesian algorithm for cross-matching multiple catalogues}",
      journal = {\mnras},
         year = 2018,
        month = feb,
       volume = {473},
       number = {4},
        pages = {4937-4955},
          doi = {10.1093/mnras/stx2651},
archivePrefix = {arXiv},
       eprint = {1705.10711},
 primaryClass = {astro-ph.GA},
       adsurl = {https://ui.adsabs.harvard.edu/abs/2018MNRAS.473.4937S}
}

@ARTICLE{Dwelly2017,
       author = {{Dwelly}, T. and {Salvato}, M. and {Merloni}, A. and {Brusa}, M. and {Buchner}, J. and {Anderson}, S.~F. and {Boller}, Th. and {Brandt}, W.~N. and {Budav{\'a}ri}, T. and {Clerc}, N. and {Coffey}, D. and {Del Moro}, A. and {Georgakakis}, A. and {Green}, P.~J. and {Jin}, C. and {Menzel}, M. -L. and {Myers}, A.~D. and {Nandra}, K. and {Nichol}, R.~C. and {Ridl}, J. and {Schwope}, A.~D. and {Simm}, T.},
        title = "{SPIDERS: selection of spectroscopic targets using AGN candidates detected in all-sky X-ray surveys}",
      journal = {\mnras},
         year = 2017,
        month = jul,
       volume = {469},
       number = {1},
        pages = {1065-1095},
          doi = {10.1093/mnras/stx864},
archivePrefix = {arXiv},
       eprint = {1704.01796},
 primaryClass = {astro-ph.GA},
       adsurl = {https://ui.adsabs.harvard.edu/abs/2017MNRAS.469.1065D}
}

@ARTICLE{Capaccioli2011,
       author = {{Capaccioli}, M. and {Schipani}, P.},
        title = "{The VLT Survey Telescope Opens to the Sky: History of a Commissioning}",
      journal = {The Messenger},
         year = 2011,
        month = dec,
       volume = {146},
        pages = {2-7},
       adsurl = {https://ui.adsabs.harvard.edu/abs/2011Msngr.146....2C}
}

@ARTICLE{Gezari2013,
       author = {{Gezari}, S. and {Martin}, D.~C. and {Forster}, K. and {Neill}, J.~D. and {Huber}, M. and {Heckman}, T. and {Bianchi}, L. and {Morrissey}, P. and {Neff}, S.~G. and {Seibert}, M. and {Schiminovich}, D. and {Wyder}, T.~K. and {Burgett}, W.~S. and {Chambers}, K.~C. and {Kaiser}, N. and {Magnier}, E.~A. and {Price}, P.~A. and {Tonry}, J.~L.},
        title = "{The GALEX Time Domain Survey. I. Selection and Classification of Over a Thousand Ultraviolet Variable Sources}",
      journal = {\apj},
         year = 2013,
        month = mar,
       volume = {766},
       number = {1},
          eid = {60},
        pages = {60},
          doi = {10.1088/0004-637X/766/1/60},
archivePrefix = {arXiv},
       eprint = {1302.1581},
 primaryClass = {astro-ph.CO},
       adsurl = {https://ui.adsabs.harvard.edu/abs/2013ApJ...766...60G}
}

@ARTICLE{Wasleske2022,
       author = {{Wasleske}, Erik J. and {Baldassare}, Vivienne F. and {Carroll}, Christopher M.},
        title = "{Variable Active Galactic Nuclei in the Galaxy Evolution Explorer Time Domain Survey}",
      journal = {\apj},
         year = 2022,
        month = jul,
       volume = {933},
       number = {1},
          eid = {37},
        pages = {37},
          doi = {10.3847/1538-4357/ac715b},
       adsurl = {https://ui.adsabs.harvard.edu/abs/2022ApJ...933...37W}
}

@article{FOLGADO2018268,
title = {Time Alignment Measurement for Time Series},
journal = {Pattern Recognition},
volume = {81},
pages = {268-279},
year = {2018},
issn = {0031-3203},
doi = {https://doi.org/10.1016/j.patcog.2018.04.003},
url = {https://www.sciencedirect.com/science/article/pii/S0031320318301286},
author = {Duarte Folgado and Marília Barandas and Ricardo Matias and Rodrigo Martins and Miguel Carvalho and Hugo Gamboa}
}

@inproceedings{Williams1995,
 author = {Williams, Christopher and Rasmussen, Carl},
 booktitle = {Advances in Neural Information Processing Systems},
 editor = {D. Touretzky and M.C. Mozer and M. Hasselmo},
 pages = {},
 publisher = {MIT Press},
 title = {Gaussian Processes for Regression},
 url = {https://proceedings.neurips.cc/paper_files/paper/1995/file/7cce53cf90577442771720a370c3c723-Paper.pdf},
 volume = {8},
 year = {1995}
}

@ARTICLE{Gorski2005,
       author = {{G{\'o}rski}, K.~M. and {Hivon}, E. and {Banday}, A.~J. and {Wandelt}, B.~D. and {Hansen}, F.~K. and {Reinecke}, M. and {Bartelmann}, M.},
        title = "{HEALPix: A Framework for High-Resolution Discretization and Fast Analysis of Data Distributed on the Sphere}",
      journal = {\apj},
         year = 2005,
        month = apr,
       volume = {622},
       number = {2},
        pages = {759-771},
          doi = {10.1086/427976},
archivePrefix = {arXiv},
       eprint = {astro-ph/0409513},
 primaryClass = {astro-ph},
       adsurl = {https://ui.adsabs.harvard.edu/abs/2005ApJ...622..759G}
}

@ARTICLE{Gaia2023,
       author = {{Gaia Collaboration} and {Vallenari}, A. and {Brown}, A.~G.~A. and {Prusti}, T. and {de Bruijne}, J.~H.~J. and {Arenou}, F. and {Babusiaux}, C. and {Biermann}, M. and {Creevey}, O.~L. and {Ducourant}, C. and {Evans}, D.~W. and {Eyer}, L. and {Guerra}, R. and {Hutton}, A. and {Jordi}, C. and {Klioner}, S.~A. and {Lammers}, U.~L. and {Lindegren}, L. and {Luri}, X. and {Mignard}, F. and {Panem}, C. and {Pourbaix}, D. and {Randich}, S. and {Sartoretti}, P. and {Soubiran}, C. and {Tanga}, P. and {Walton}, N.~A. and {Bailer-Jones}, C.~A.~L. and {Bastian}, U. and {Drimmel}, R. and {Jansen}, F. and {Katz}, D. and {Lattanzi}, M.~G. and {van Leeuwen}, F. and {Bakker}, J. and {Cacciari}, C. and {Casta{\~n}eda}, J. and {De Angeli}, F. and {Fabricius}, C. and {Fouesneau}, M. and {Fr{\'e}mat}, Y. and {Galluccio}, L. and {Guerrier}, A. and {Heiter}, U. and {Masana}, E. and {Messineo}, R. and {Mowlavi}, N. and {Nicolas}, C. and {Nienartowicz}, K. and {Pailler}, F. and {Panuzzo}, P. and {Riclet}, F. and {Roux}, W. and {Seabroke}, G.~M. and {Sordo}, R. and {Th{\'e}venin}, F. and {Gracia-Abril}, G. and {Portell}, J. and {Teyssier}, D. and {Altmann}, M. and {Andrae}, R. and {Audard}, M. and {Bellas-Velidis}, I. and {Benson}, K. and {Berthier}, J. and {Blomme}, R. and {Burgess}, P.~W. and {Busonero}, D. and {Busso}, G. and {C{\'a}novas}, H. and {Carry}, B. and {Cellino}, A. and {Cheek}, N. and {Clementini}, G. and {Damerdji}, Y. and {Davidson}, M. and {de Teodoro}, P. and {Nu{\~n}ez Campos}, M. and {Delchambre}, L. and {Dell'Oro}, A. and {Esquej}, P. and {Fern{\'a}ndez-Hern{\'a}ndez}, J. and {Fraile}, E. and {Garabato}, D. and {Garc{\'\i}a-Lario}, P. and {Gosset}, E. and {Haigron}, R. and {Halbwachs}, J. -L. and {Hambly}, N.~C. and {Harrison}, D.~L. and {Hern{\'a}ndez}, J. and {Hestroffer}, D. and {Hodgkin}, S.~T. and {Holl}, B. and {Jan{\ss}en}, K. and {Jevardat de Fombelle}, G. and {Jordan}, S. and {Krone-Martins}, A. and {Lanzafame}, A.~C. and {L{\"o}ffler}, W. and {Marchal}, O. and {Marrese}, P.~M. and {Moitinho}, A. and {Muinonen}, K. and {Osborne}, P. and {Pancino}, E. and {Pauwels}, T. and {Recio-Blanco}, A. and {Reyl{\'e}}, C. and {Riello}, M. and {Rimoldini}, L. and {Roegiers}, T. and {Rybizki}, J. and {Sarro}, L.~M. and {Siopis}, C. and {Smith}, M. and {Sozzetti}, A. and {Utrilla}, E. and {van Leeuwen}, M. and {Abbas}, U. and {{\'A}brah{\'a}m}, P. and {Abreu Aramburu}, A. and {Aerts}, C. and {Aguado}, J.~J. and {Ajaj}, M. and {Aldea-Montero}, F. and {Altavilla}, G. and {{\'A}lvarez}, M.~A. and {Alves}, J. and {Anders}, F. and {Anderson}, R.~I. and {Anglada Varela}, E. and {Antoja}, T. and {Baines}, D. and {Baker}, S.~G. and {Balaguer-N{\'u}{\~n}ez}, L. and {Balbinot}, E. and {Balog}, Z. and {Barache}, C. and {Barbato}, D. and {Barros}, M. and {Barstow}, M.~A. and {Bartolom{\'e}}, S. and {Bassilana}, J. -L. and {Bauchet}, N. and {Becciani}, U. and {Bellazzini}, M. and {Berihuete}, A. and {Bernet}, M. and {Bertone}, S. and {Bianchi}, L. and {Binnenfeld}, A. and {Blanco-Cuaresma}, S. and {Blazere}, A. and {Boch}, T. and {Bombrun}, A. and {Bossini}, D. and {Bouquillon}, S. and {Bragaglia}, A. and {Bramante}, L. and {Breedt}, E. and {Bressan}, A. and {Brouillet}, N. and {Brugaletta}, E. and {Bucciarelli}, B. and {Burlacu}, A. and {Butkevich}, A.~G. and {Buzzi}, R. and {Caffau}, E. and {Cancelliere}, R. and {Cantat-Gaudin}, T. and {Carballo}, R. and {Carlucci}, T. and {Carnerero}, M.~I. and {Carrasco}, J.~M. and {Casamiquela}, L. and {Castellani}, M. and {Castro-Ginard}, A. and {Chaoul}, L. and {Charlot}, P. and {Chemin}, L. and {Chiaramida}, V. and {Chiavassa}, A. and {Chornay}, N. and {Comoretto}, G. and {Contursi}, G. and {Cooper}, W.~J. and {Cornez}, T. and {Cowell}, S. and {Crifo}, F. and {Cropper}, M. and {Crosta}, M. and {Crowley}, C. and {Dafonte}, C. and {Dapergolas}, A. and {David}, M. and {David}, P. and {de Laverny}, P. and {De Luise}, F. and {De March}, R. and {De Ridder}, J. and {de Souza}, R. and {de Torres}, A. and {del Peloso}, E.~F. and {del Pozo}, E. and {Delbo}, M. and {Delgado}, A. and {Delisle}, J. -B. and {Demouchy}, C. and {Dharmawardena}, T.~E. and {Di Matteo}, P. and {Diakite}, S. and {Diener}, C. and {Distefano}, E. and {Dolding}, C. and {Edvardsson}, B. and {Enke}, H. and {Fabre}, C. and {Fabrizio}, M. and {Faigler}, S. and {Fedorets}, G. and {Fernique}, P. and {Fienga}, A. and {Figueras}, F. and {Fournier}, Y. and {Fouron}, C. and {Fragkoudi}, F. and {Gai}, M. and {Garcia-Gutierrez}, A. and {Garcia-Reinaldos}, M. and {Garc{\'\i}a-Torres}, M. and {Garofalo}, A. and {Gavel}, A. and {Gavras}, P. and {Gerlach}, E. and {Geyer}, R. and {Giacobbe}, P. and {Gilmore}, G. and {Girona}, S. and {Giuffrida}, G. and {Gomel}, R. and {Gomez}, A. and {Gonz{\'a}lez-N{\'u}{\~n}ez}, J. and {Gonz{\'a}lez-Santamar{\'\i}a}, I. and {Gonz{\'a}lez-Vidal}, J.~J. and {Granvik}, M. and {Guillout}, P. and {Guiraud}, J. and {Guti{\'e}rrez-S{\'a}nchez}, R. and {Guy}, L.~P. and {Hatzidimitriou}, D. and {Hauser}, M. and {Haywood}, M. and {Helmer}, A. and {Helmi}, A. and {Sarmiento}, M.~H. and {Hidalgo}, S.~L. and {Hilger}, T. and {H{\l}adczuk}, N. and {Hobbs}, D. and {Holland}, G. and {Huckle}, H.~E. and {Jardine}, K. and {Jasniewicz}, G. and {Jean-Antoine Piccolo}, A. and {Jim{\'e}nez-Arranz}, {\'O}. and {Jorissen}, A. and {Juaristi Campillo}, J. and {Julbe}, F. and {Karbevska}, L. and {Kervella}, P. and {Khanna}, S. and {Kontizas}, M. and {Kordopatis}, G. and {Korn}, A.~J. and {K{\'o}sp{\'a}l}, {\'A}. and {Kostrzewa-Rutkowska}, Z. and {Kruszy{\'n}ska}, K. and {Kun}, M. and {Laizeau}, P. and {Lambert}, S. and {Lanza}, A.~F. and {Lasne}, Y. and {Le Campion}, J. -F. and {Lebreton}, Y. and {Lebzelter}, T. and {Leccia}, S. and {Leclerc}, N. and {Lecoeur-Taibi}, I. and {Liao}, S. and {Licata}, E.~L. and {Lindstr{\o}m}, H.~E.~P. and {Lister}, T.~A. and {Livanou}, E. and {Lobel}, A. and {Lorca}, A. and {Loup}, C. and {Madrero Pardo}, P. and {Magdaleno Romeo}, A. and {Managau}, S. and {Mann}, R.~G. and {Manteiga}, M. and {Marchant}, J.~M. and {Marconi}, M. and {Marcos}, J. and {Marcos Santos}, M.~M.~S. and {Mar{\'\i}n Pina}, D. and {Marinoni}, S. and {Marocco}, F. and {Marshall}, D.~J. and {Martin Polo}, L. and {Mart{\'\i}n-Fleitas}, J.~M. and {Marton}, G. and {Mary}, N. and {Masip}, A. and {Massari}, D. and {Mastrobuono-Battisti}, A. and {Mazeh}, T. and {McMillan}, P.~J. and {Messina}, S. and {Michalik}, D. and {Millar}, N.~R. and {Mints}, A. and {Molina}, D. and {Molinaro}, R. and {Moln{\'a}r}, L. and {Monari}, G. and {Mongui{\'o}}, M. and {Montegriffo}, P. and {Montero}, A. and {Mor}, R. and {Mora}, A. and {Morbidelli}, R. and {Morel}, T. and {Morris}, D. and {Muraveva}, T. and {Murphy}, C.~P. and {Musella}, I. and {Nagy}, Z. and {Noval}, L. and {Oca{\~n}a}, F. and {Ogden}, A. and {Ordenovic}, C. and {Osinde}, J.~O. and {Pagani}, C. and {Pagano}, I. and {Palaversa}, L. and {Palicio}, P.~A. and {Pallas-Quintela}, L. and {Panahi}, A. and {Payne-Wardenaar}, S. and {Pe{\~n}alosa Esteller}, X. and {Penttil{\"a}}, A. and {Pichon}, B. and {Piersimoni}, A.~M. and {Pineau}, F. -X. and {Plachy}, E. and {Plum}, G. and {Poggio}, E. and {Pr{\v{s}}a}, A. and {Pulone}, L. and {Racero}, E. and {Ragaini}, S. and {Rainer}, M. and {Raiteri}, C.~M. and {Rambaux}, N. and {Ramos}, P. and {Ramos-Lerate}, M. and {Re Fiorentin}, P. and {Regibo}, S. and {Richards}, P.~J. and {Rios Diaz}, C. and {Ripepi}, V. and {Riva}, A. and {Rix}, H. -W. and {Rixon}, G. and {Robichon}, N. and {Robin}, A.~C. and {Robin}, C. and {Roelens}, M. and {Rogues}, H.~R.~O. and {Rohrbasser}, L. and {Romero-G{\'o}mez}, M. and {Rowell}, N. and {Royer}, F. and {Ruz Mieres}, D. and {Rybicki}, K.~A. and {Sadowski}, G. and {S{\'a}ez N{\'u}{\~n}ez}, A. and {Sagrist{\`a} Sell{\'e}s}, A. and {Sahlmann}, J. and {Salguero}, E. and {Samaras}, N. and {Sanchez Gimenez}, V. and {Sanna}, N. and {Santove{\~n}a}, R. and {Sarasso}, M. and {Schultheis}, M. and {Sciacca}, E. and {Segol}, M. and {Segovia}, J.~C. and {S{\'e}gransan}, D. and {Semeux}, D. and {Shahaf}, S. and {Siddiqui}, H.~I. and {Siebert}, A. and {Siltala}, L. and {Silvelo}, A. and {Slezak}, E. and {Slezak}, I. and {Smart}, R.~L. and {Snaith}, O.~N. and {Solano}, E. and {Solitro}, F. and {Souami}, D. and {Souchay}, J. and {Spagna}, A. and {Spina}, L. and {Spoto}, F. and {Steele}, I.~A. and {Steidelm{\"u}ller}, H. and {Stephenson}, C.~A. and {S{\"u}veges}, M. and {Surdej}, J. and {Szabados}, L. and {Szegedi-Elek}, E. and {Taris}, F. and {Taylor}, M.~B. and {Teixeira}, R. and {Tolomei}, L. and {Tonello}, N. and {Torra}, F. and {Torra}, J. and {Torralba Elipe}, G. and {Trabucchi}, M. and {Tsounis}, A.~T. and {Turon}, C. and {Ulla}, A. and {Unger}, N. and {Vaillant}, M.~V. and {van Dillen}, E. and {van Reeven}, W. and {Vanel}, O. and {Vecchiato}, A. and {Viala}, Y. and {Vicente}, D. and {Voutsinas}, S. and {Weiler}, M. and {Wevers}, T. and {Wyrzykowski}, {\L}. and {Yoldas}, A. and {Yvard}, P. and {Zhao}, H. and {Zorec}, J. and {Zucker}, S. and {Zwitter}, T.},
        title = "{Gaia Data Release 3. Summary of the content and survey properties}",
      journal = {\aap},
         year = 2023,
        month = jun,
       volume = {674},
          eid = {A1},
        pages = {A1},
          doi = {10.1051/0004-6361/202243940},
archivePrefix = {arXiv},
       eprint = {2208.00211},
 primaryClass = {astro-ph.GA},
       adsurl = {https://ui.adsabs.harvard.edu/abs/2023A&A...674A...1G}
}

@ARTICLE{Flewelling2020,
       author = {{Flewelling}, H.~A. and {Magnier}, E.~A. and {Chambers}, K.~C. and {Heasley}, J.~N. and {Holmberg}, C. and {Huber}, M.~E. and {Sweeney}, W. and {Waters}, C.~Z. and {Calamida}, A. and {Casertano}, S. and {Chen}, X. and {Farrow}, D. and {Hasinger}, G. and {Henderson}, R. and {Long}, K.~S. and {Metcalfe}, N. and {Narayan}, G. and {Nieto-Santisteban}, M.~A. and {Norberg}, P. and {Rest}, A. and {Saglia}, R.~P. and {Szalay}, A. and {Thakar}, A.~R. and {Tonry}, J.~L. and {Valenti}, J. and {Werner}, S. and {White}, R. and {Denneau}, L. and {Draper}, P.~W. and {Hodapp}, K.~W. and {Jedicke}, R. and {Kaiser}, N. and {Kudritzki}, R.~P. and {Price}, P.~A. and {Wainscoat}, R.~J. and {Chastel}, S. and {McLean}, B. and {Postman}, M. and {Shiao}, B.},
        title = "{The Pan-STARRS1 Database and Data Products}",
      journal = {\apjs},
         year = 2020,
        month = nov,
       volume = {251},
       number = {1},
          eid = {7},
        pages = {7},
          doi = {10.3847/1538-4365/abb82d},
archivePrefix = {arXiv},
       eprint = {1612.05243},
 primaryClass = {astro-ph.IM},
       adsurl = {https://ui.adsabs.harvard.edu/abs/2020ApJS..251....7F}
}

@ARTICLE{Mainzer2011,
       author = {{Mainzer}, A. and {Bauer}, J. and {Grav}, T. and {Masiero}, J. and {Cutri}, R.~M. and {Dailey}, J. and {Eisenhardt}, P. and {McMillan}, R.~S. and {Wright}, E. and {Walker}, R. and {Jedicke}, R. and {Spahr}, T. and {Tholen}, D. and {Alles}, R. and {Beck}, R. and {Brandenburg}, H. and {Conrow}, T. and {Evans}, T. and {Fowler}, J. and {Jarrett}, T. and {Marsh}, K. and {Masci}, F. and {McCallon}, H. and {Wheelock}, S. and {Wittman}, M. and {Wyatt}, P. and {DeBaun}, E. and {Elliott}, G. and {Elsbury}, D. and {Gautier}, T., IV and {Gomillion}, S. and {Leisawitz}, D. and {Maleszewski}, C. and {Micheli}, M. and {Wilkins}, A.},
        title = "{Preliminary Results from NEOWISE: An Enhancement to the Wide-field Infrared Survey Explorer for Solar System Science}",
      journal = {\apj},
         year = 2011,
        month = apr,
       volume = {731},
       number = {1},
          eid = {53},
        pages = {53},
          doi = {10.1088/0004-637X/731/1/53},
archivePrefix = {arXiv},
       eprint = {1102.1996},
 primaryClass = {astro-ph.EP},
       adsurl = {https://ui.adsabs.harvard.edu/abs/2011ApJ...731...53M}
}

@ARTICLE{Meisner2023,
       author = {{Meisner}, Aaron M. and {Caselden}, Dan and {Schlafly}, Edward F. and {Kiwy}, Frank},
        title = "{unTimely: a Full-sky, Time-domain unWISE Catalog}",
      journal = {\aj},
         year = 2023,
        month = feb,
       volume = {165},
       number = {2},
          eid = {36},
        pages = {36},
          doi = {10.3847/1538-3881/aca2ab},
archivePrefix = {arXiv},
       eprint = {2209.14327},
 primaryClass = {astro-ph.IM},
       adsurl = {https://ui.adsabs.harvard.edu/abs/2023AJ....165...36M}
}

@ARTICLE{Masci2019,
       author = {{Masci}, Frank J. and {Laher}, Russ R. and {Rusholme}, Ben and {Shupe}, David L. and {Groom}, Steven and {Surace}, Jason and {Jackson}, Edward and {Monkewitz}, Serge and {Beck}, Ron and {Flynn}, David and {Terek}, Scott and {Landry}, Walter and {Hacopians}, Eugean and {Desai}, Vandana and {Howell}, Justin and {Brooke}, Tim and {Imel}, David and {Wachter}, Stefanie and {Ye}, Quan-Zhi and {Lin}, Hsing-Wen and {Cenko}, S. Bradley and {Cunningham}, Virginia and {Rebbapragada}, Umaa and {Bue}, Brian and {Miller}, Adam A. and {Mahabal}, Ashish and {Bellm}, Eric C. and {Patterson}, Maria T. and {Juri{\'c}}, Mario and {Golkhou}, V. Zach and {Ofek}, Eran O. and {Walters}, Richard and {Graham}, Matthew and {Kasliwal}, Mansi M. and {Dekany}, Richard G. and {Kupfer}, Thomas and {Burdge}, Kevin and {Cannella}, Christopher B. and {Barlow}, Tom and {Van Sistine}, Angela and {Giomi}, Matteo and {Fremling}, Christoffer and {Blagorodnova}, Nadejda and {Levitan}, David and {Riddle}, Reed and {Smith}, Roger M. and {Helou}, George and {Prince}, Thomas A. and {Kulkarni}, Shrinivas R.},
        title = "{The Zwicky Transient Facility: Data Processing, Products, and Archive}",
      journal = {\pasp},
         year = 2019,
        month = jan,
       volume = {131},
       number = {995},
        pages = {018003},
          doi = {10.1088/1538-3873/aae8ac},
archivePrefix = {arXiv},
       eprint = {1902.01872},
 primaryClass = {astro-ph.IM},
       adsurl = {https://ui.adsabs.harvard.edu/abs/2019PASP..131a8003M}
}

@ARTICLE{Pan-STARRS,
       author = {{Chambers}, K.~C. and {Magnier}, E.~A. and {Metcalfe}, N. and {Flewelling}, H.~A. and {Huber}, M.~E. and {Waters}, C.~Z. and {Denneau}, L. and {Draper}, P.~W. and {Farrow}, D. and {Finkbeiner}, D.~P. and {Holmberg}, C. and {Koppenhoefer}, J. and {Price}, P.~A. and {Rest}, A. and {Saglia}, R.~P. and {Schlafly}, E.~F. and {Smartt}, S.~J. and {Sweeney}, W. and {Wainscoat}, R.~J. and {Burgett}, W.~S. and {Chastel}, S. and {Grav}, T. and {Heasley}, J.~N. and {Hodapp}, K.~W. and {Jedicke}, R. and {Kaiser}, N. and {Kudritzki}, R. -P. and {Luppino}, G.~A. and {Lupton}, R.~H. and {Monet}, D.~G. and {Morgan}, J.~S. and {Onaka}, P.~M. and {Shiao}, B. and {Stubbs}, C.~W. and {Tonry}, J.~L. and {White}, R. and {Ba{\~n}ados}, E. and {Bell}, E.~F. and {Bender}, R. and {Bernard}, E.~J. and {Boegner}, M. and {Boffi}, F. and {Botticella}, M.~T. and {Calamida}, A. and {Casertano}, S. and {Chen}, W. -P. and {Chen}, X. and {Cole}, S. and {Deacon}, N. and {Frenk}, C. and {Fitzsimmons}, A. and {Gezari}, S. and {Gibbs}, V. and {Goessl}, C. and {Goggia}, T. and {Gourgue}, R. and {Goldman}, B. and {Grant}, P. and {Grebel}, E.~K. and {Hambly}, N.~C. and {Hasinger}, G. and {Heavens}, A.~F. and {Heckman}, T.~M. and {Henderson}, R. and {Henning}, T. and {Holman}, M. and {Hopp}, U. and {Ip}, W. -H. and {Isani}, S. and {Jackson}, M. and {Keyes}, C.~D. and {Koekemoer}, A.~M. and {Kotak}, R. and {Le}, D. and {Liska}, D. and {Long}, K.~S. and {Lucey}, J.~R. and {Liu}, M. and {Martin}, N.~F. and {Masci}, G. and {McLean}, B. and {Mindel}, E. and {Misra}, P. and {Morganson}, E. and {Murphy}, D.~N.~A. and {Obaika}, A. and {Narayan}, G. and {Nieto-Santisteban}, M.~A. and {Norberg}, P. and {Peacock}, J.~A. and {Pier}, E.~A. and {Postman}, M. and {Primak}, N. and {Rae}, C. and {Rai}, A. and {Riess}, A. and {Riffeser}, A. and {Rix}, H.~W. and {R{\"o}ser}, S. and {Russel}, R. and {Rutz}, L. and {Schilbach}, E. and {Schultz}, A.~S.~B. and {Scolnic}, D. and {Strolger}, L. and {Szalay}, A. and {Seitz}, S. and {Small}, E. and {Smith}, K.~W. and {Soderblom}, D.~R. and {Taylor}, P. and {Thomson}, R. and {Taylor}, A.~N. and {Thakar}, A.~R. and {Thiel}, J. and {Thilker}, D. and {Unger}, D. and {Urata}, Y. and {Valenti}, J. and {Wagner}, J. and {Walder}, T. and {Walter}, F. and {Watters}, S.~P. and {Werner}, S. and {Wood-Vasey}, W.~M. and {Wyse}, R.},
        title = "{The Pan-STARRS1 Surveys}",
      journal = {arXiv e-prints},
         year = 2016,
        month = dec,
          eid = {arXiv:1612.05560},
        pages = {arXiv:1612.05560},
          doi = {10.48550/arXiv.1612.05560},
archivePrefix = {arXiv},
       eprint = {1612.05560},
 primaryClass = {astro-ph.IM},
       adsurl = {https://ui.adsabs.harvard.edu/abs/2016arXiv161205560C}
}

@ARTICLE{ZTF,
       author = {{Bellm}, Eric C. and {Kulkarni}, Shrinivas R. and {Graham}, Matthew J. and {Dekany}, Richard and {Smith}, Roger M. and {Riddle}, Reed and {Masci}, Frank J. and {Helou}, George and {Prince}, Thomas A. and {Adams}, Scott M. and {Barbarino}, C. and {Barlow}, Tom and {Bauer}, James and {Beck}, Ron and {Belicki}, Justin and {Biswas}, Rahul and {Blagorodnova}, Nadejda and {Bodewits}, Dennis and {Bolin}, Bryce and {Brinnel}, Valery and {Brooke}, Tim and {Bue}, Brian and {Bulla}, Mattia and {Burruss}, Rick and {Cenko}, S. Bradley and {Chang}, Chan-Kao and {Connolly}, Andrew and {Coughlin}, Michael and {Cromer}, John and {Cunningham}, Virginia and {De}, Kishalay and {Delacroix}, Alex and {Desai}, Vandana and {Duev}, Dmitry A. and {Eadie}, Gwendolyn and {Farnham}, Tony L. and {Feeney}, Michael and {Feindt}, Ulrich and {Flynn}, David and {Franckowiak}, Anna and {Frederick}, S. and {Fremling}, C. and {Gal-Yam}, Avishay and {Gezari}, Suvi and {Giomi}, Matteo and {Goldstein}, Daniel A. and {Golkhou}, V. Zach and {Goobar}, Ariel and {Groom}, Steven and {Hacopians}, Eugean and {Hale}, David and {Henning}, John and {Ho}, Anna Y.~Q. and {Hover}, David and {Howell}, Justin and {Hung}, Tiara and {Huppenkothen}, Daniela and {Imel}, David and {Ip}, Wing-Huen and {Ivezi{\'c}}, {\v{Z}}eljko and {Jackson}, Edward and {Jones}, Lynne and {Juric}, Mario and {Kasliwal}, Mansi M. and {Kaspi}, S. and {Kaye}, Stephen and {Kelley}, Michael S.~P. and {Kowalski}, Marek and {Kramer}, Emily and {Kupfer}, Thomas and {Landry}, Walter and {Laher}, Russ R. and {Lee}, Chien-De and {Lin}, Hsing Wen and {Lin}, Zhong-Yi and {Lunnan}, Ragnhild and {Giomi}, Matteo and {Mahabal}, Ashish and {Mao}, Peter and {Miller}, Adam A. and {Monkewitz}, Serge and {Murphy}, Patrick and {Ngeow}, Chow-Choong and {Nordin}, Jakob and {Nugent}, Peter and {Ofek}, Eran and {Patterson}, Maria T. and {Penprase}, Bryan and {Porter}, Michael and {Rauch}, Ludwig and {Rebbapragada}, Umaa and {Reiley}, Dan and {Rigault}, Mickael and {Rodriguez}, Hector and {van Roestel}, Jan and {Rusholme}, Ben and {van Santen}, Jakob and {Schulze}, S. and {Shupe}, David L. and {Singer}, Leo P. and {Soumagnac}, Maayane T. and {Stein}, Robert and {Surace}, Jason and {Sollerman}, Jesper and {Szkody}, Paula and {Taddia}, F. and {Terek}, Scott and {Van Sistine}, Angela and {van Velzen}, Sjoert and {Vestrand}, W. Thomas and {Walters}, Richard and {Ward}, Charlotte and {Ye}, Quan-Zhi and {Yu}, Po-Chieh and {Yan}, Lin and {Zolkower}, Jeffry},
        title = "{The Zwicky Transient Facility: System Overview, Performance, and First Results}",
      journal = {\pasp},
         year = 2019,
        month = jan,
       volume = {131},
       number = {995},
        pages = {018002},
          doi = {10.1088/1538-3873/aaecbe},
archivePrefix = {arXiv},
       eprint = {1902.01932},
 primaryClass = {astro-ph.IM},
       adsurl = {https://ui.adsabs.harvard.edu/abs/2019PASP..131a8002B}
}

@ARTICLE{WISE,
       author = {{Wright}, Edward L. and {Eisenhardt}, Peter R.~M. and {Mainzer}, Amy K. and {Ressler}, Michael E. and {Cutri}, Roc M. and {Jarrett}, Thomas and {Kirkpatrick}, J. Davy and {Padgett}, Deborah and {McMillan}, Robert S. and {Skrutskie}, Michael and {Stanford}, S.~A. and {Cohen}, Martin and {Walker}, Russell G. and {Mather}, John C. and {Leisawitz}, David and {Gautier}, Thomas N., III and {McLean}, Ian and {Benford}, Dominic and {Lonsdale}, Carol J. and {Blain}, Andrew and {Mendez}, Bryan and {Irace}, William R. and {Duval}, Valerie and {Liu}, Fengchuan and {Royer}, Don and {Heinrichsen}, Ingolf and {Howard}, Joan and {Shannon}, Mark and {Kendall}, Martha and {Walsh}, Amy L. and {Larsen}, Mark and {Cardon}, Joel G. and {Schick}, Scott and {Schwalm}, Mark and {Abid}, Mohamed and {Fabinsky}, Beth and {Naes}, Larry and {Tsai}, Chao-Wei},
        title = "{The Wide-field Infrared Survey Explorer (WISE): Mission Description and Initial On-orbit Performance}",
      journal = {\aj},
         year = 2010,
        month = dec,
       volume = {140},
       number = {6},
        pages = {1868-1881},
          doi = {10.1088/0004-6256/140/6/1868},
archivePrefix = {arXiv},
       eprint = {1008.0031},
 primaryClass = {astro-ph.IM},
       adsurl = {https://ui.adsabs.harvard.edu/abs/2010AJ....140.1868W}
}

@article{scikit-learn,
  title={Scikit-learn: Machine Learning in {P}ython},
  author={Pedregosa, F. and Varoquaux, G. and Gramfort, A. and Michel, V.
          and Thirion, B. and Grisel, O. and Blondel, M. and Prettenhofer, P.
          and Weiss, R. and Dubourg, V. and Vanderplas, J. and Passos, A. and
          Cournapeau, D. and Brucher, M. and Perrot, M. and Duchesnay, E.},
  journal={Journal of Machine Learning Research},
  volume={12},
  pages={2825--2830},
  year={2011}
}

@ARTICLE{McInnes2018,
       author = {{McInnes}, Leland and {Healy}, John and {Melville}, James},
        title = "{UMAP: Uniform Manifold Approximation and Projection for Dimension Reduction}",
      journal = {arXiv e-prints},
         year = 2018,
        month = feb,
          eid = {arXiv:1802.03426},
        pages = {arXiv:1802.03426},
          doi = {10.48550/arXiv.1802.03426},
archivePrefix = {arXiv},
       eprint = {1802.03426},
 primaryClass = {stat.ML},
       adsurl = {https://ui.adsabs.harvard.edu/abs/2018arXiv180203426M}
}

@ARTICLE{Lyke2020,
       author = {{Lyke}, Brad W. and {Higley}, Alexandra N. and {McLane}, J.~N. and {Schurhammer}, Danielle P. and {Myers}, Adam D. and {Ross}, Ashley J. and {Dawson}, Kyle and {Chabanier}, Sol{\`e}ne and {Martini}, Paul and {Busca}, Nicol{\'a}s G. and {Mas des Bourboux}, H{\'e}lion du and {Salvato}, Mara and {Streblyanska}, Alina and {Zarrouk}, Pauline and {Burtin}, Etienne and {Anderson}, Scott F. and {Bautista}, Julian and {Bizyaev}, Dmitry and {Brandt}, W.~N. and {Brinkmann}, Jonathan and {Brownstein}, Joel R. and {Comparat}, Johan and {Green}, Paul and {de la Macorra}, Axel and {Mu{\~n}oz Guti{\'e}rrez}, Andrea and {Hou}, Jiamin and {Newman}, Jeffrey A. and {Palanque-Delabrouille}, Nathalie and {P{\^a}ris}, Isabelle and {Percival}, Will J. and {Petitjean}, Patrick and {Rich}, James and {Rossi}, Graziano and {Schneider}, Donald P. and {Smith}, Alexander and {Vivek}, M. and {Weaver}, Benjamin Alan},
        title = "{The Sloan Digital Sky Survey Quasar Catalog: Sixteenth Data Release}",
      journal = {\apjs},
         year = 2020,
        month = sep,
       volume = {250},
       number = {1},
          eid = {8},
        pages = {8},
          doi = {10.3847/1538-4365/aba623},
archivePrefix = {arXiv},
       eprint = {2007.09001},
 primaryClass = {astro-ph.GA},
       adsurl = {https://ui.adsabs.harvard.edu/abs/2020ApJS..250....8L}
}

\end{document}